\documentclass[11pt,twoside]{article}

\usepackage{epsfig,amsfonts,color}
\usepackage{amsmath}
\usepackage{amssymb, palatino, geometry,url}

\usepackage[colorlinks=true,linkcolor=blue,citecolor=blue,urlcolor=blue]{hyperref}

\usepackage{fancyhdr}
\newcommand{\be}{\begin{equation}}
\newcommand{\ee}{\end{equation}}
\newcommand{\bea}{\begin{eqnarray}}
\newcommand{\eea}{\end{eqnarray}}

\newcommand{\bvec}{\left(\begin{array}{c}}
\newcommand{\evec}{\end{array}\right)}
\newcommand{\bsub}{\begin{subequations}}
\newcommand{\esub}{\end{subequations}}

\usepackage{lineno}

\usepackage{mathtools}
\usepackage{graphicx}
\usepackage{amsfonts}
\usepackage{caption} 
\usepackage{subcaption} 
\usepackage{mathrsfs}
\usepackage{multirow}
\usepackage{float}
\usepackage[english]{babel}
\usepackage{amsthm}
\usepackage[utf8]{inputenc}
\usepackage{appendix}

\newtheorem{theorem}{Theorem}[section]

\newcommand{\green}[1]{\textcolor{black}{#1}}

\usepackage{natbib}
 \bibpunct[, ]{(}{)}{,}{a}{}{,}%
\begin{document}

\title{Remunerating Space-Time, Load-Shifting Flexibility\\  from Data Centers in Electricity Markets}

\author{Weiqi Zhang${}^{\ddag}$ and Victor M. Zavala${}^{\ddag}$\thanks{Corresponding Author: victor.zavala@wisc.edu}\\
  {\small ${}^\ddag$Department of Chemical and Biological Engineering}\\
 {\small \;University of Wisconsin-Madison, 1415 Engineering Dr, Madison, WI 53706, USA}}
 \date{}
\maketitle

\begin{abstract}
We study an electricity market clearing formulation that seeks to remunerate spatio-temporal, load-shifting flexibility provided by data centers (DaCes). Load-shifting flexibility is a key asset for power grid operators as they aim to integrate larger amounts of intermittent renewable power and to decarbonize the grid. Central to our study is the concept of virtual links, which provide non-physical pathways that can be used by DaCes to shift power loads (by shifting computing loads) across space and time. We use virtual links to show that the clearing formulation treats DaCes as prosumers that simultaneously request load and provide a load-shifting flexibility service. Our analysis also reveals that DaCes are remunerated for the provision of load-shifting flexibility based on nodal price differences (across space and time). We also show that DaCe flexibility helps relieve space-time price volatility and show that the clearing formulation satisfies fundamental economic properties that are expected from coordinated markets (e.g., provides a competitive equilibrium and achieves revenue adequacy and cost recovery). The concepts presented are applicable to other key market players that can offer space-time shifting flexibility such as distributed manufacturing facilities and storage systems. Case studies are presented to demonstrate these properties. 
\end{abstract}


\maketitle

\section{Introduction}

\label{sec:intro}

The power grid is undergoing major structural changes due to the adoption of large amounts of renewable power and the need to decarbonize operations. Multiple U.S. states have set ambitious renewable portfolio standards (RPS) that dictate the required level of renewable energy use in the near future, including California (50\% by 2030 according to California Public Utilities Commission \citeyear{ca_rps}), Minnesota and New York (around 25\% by 2025 and 70 \% by 2030 according to National Conference of State Legislatures \citeyear{state_rps}). A critical challenge that emerges here is the unsteady, non-dispatchable, and spatio-temporal nature of renewable power.  Enabling high penetration of renewable power requires new sources of {\em load-shifting flexibility} \citep{wierman_liu_liu_mohsenian-rad_2017}. Flexibility is a key asset in power system operations that is often harnessed from consumers via demand response and price signals \citep{dr_review}.  
 
\begin{figure}[!htb]
\centering
\includegraphics[width=0.6\textwidth]{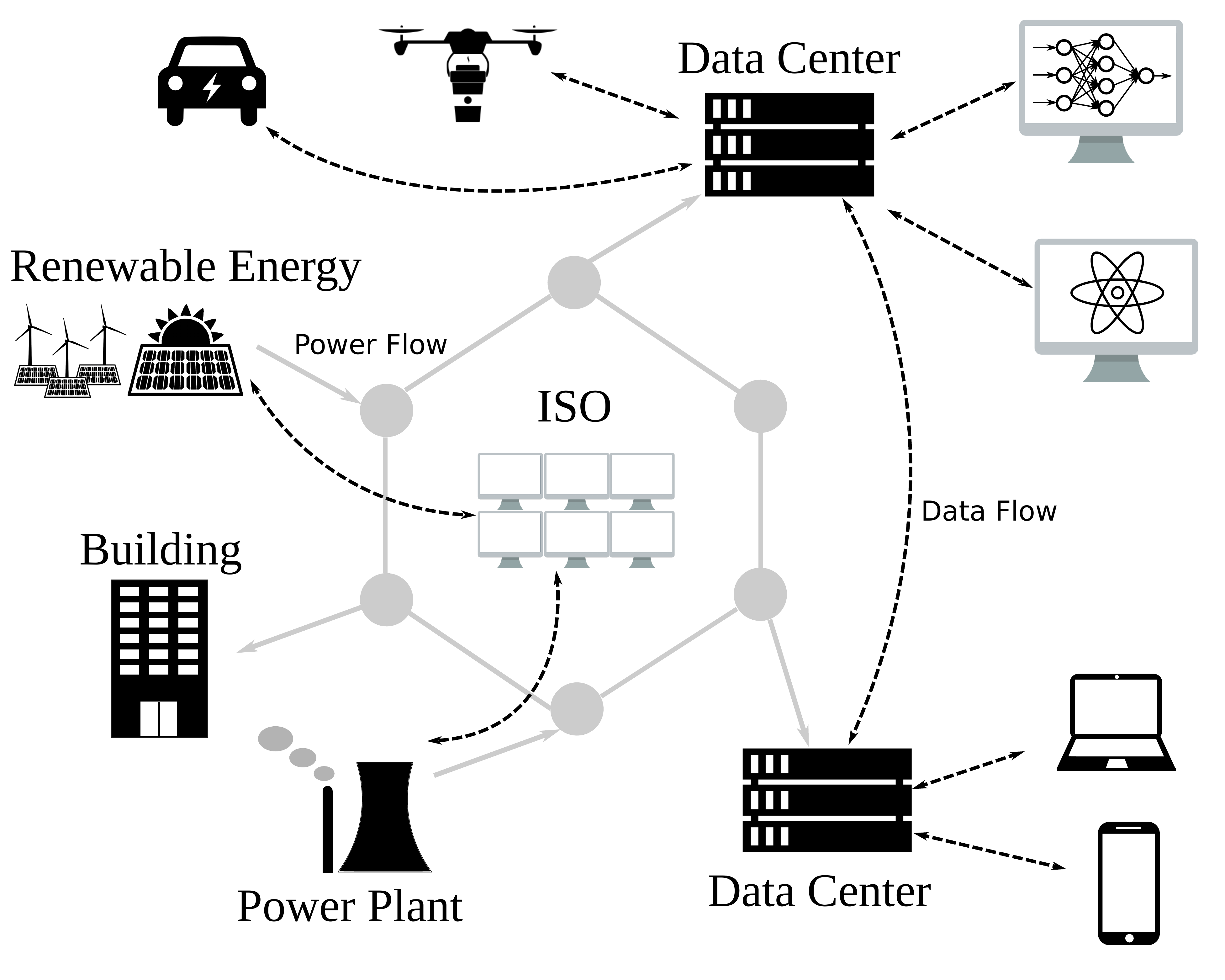}
\caption{\footnotesize Interactions between the computing and power infrastructures.}
\label{fig:infra}
\end{figure}

Rapid expansion of the computing industry also poses significant challenges to the power grid. Power use from the information technology (IT) sector is experiencing fast growth (8\% in 2016 and projected at 13\% in 2027) and the dynamics and spatial distribution of data centers (DaCes) now represents a significant demand on the grid. In addition, the computing infrastructure is undergoing structural changes; specifically, motivated by economies of scale, large companies (e.g., Amazon, Google, Alibaba, and Tencent) are centralizing DaCes. These DaCes form a computing infrastructure that is managed collectively via network operation centers (NOCs). NOCs have the ability to shift computing loads/jobs (and associated power loads) across time (via job scheduling) and across space (via service migration). As a result, NOCs can play an important role in providing {\em space-time} load-shifting flexibility to the power grid. These synergies are illustrated in Figure \ref{fig:infra}. {Exploiting load-shifting flexibility also brings important environmental benefits; for instance, Google recently introduced a Carbon-Intelligent Compute Management system that schedules flexible workloads to minimize carbon footprint (by consuming power at time or locations with low carbon content)  \citep{radovanovic2021carbon}.}

Space-time, load-shifting flexibility can also be provided by other key electricity market players such as manufacturing and storage systems. In the context of manufacturing, there is an on-going trend to deploy small-scale, modular production facilities  as a way to harness distributed and stranded resources (e.g., waste streams, biomass, and renewable power) and to gain more flexibility in both investment and operations \citep{allman2020dynamic}. The deployment of modular manufacturing systems would decentralize power loads and potentially aid power grid operations. A key example of this trend is that of ammonia and hydrogen manufacturing, which are currently produced at large centralized facilities \citep{smith2020current}. At the same time, it has been recently shown that space-time electricity market dynamics incentivize the deployment of modular systems and to decentralize loads; this is because exploiting space-time dynamics provides investors with a mechanism to mitigate risk (by exploiting price differences at across space and time \citep{shao2019space}.  

There has been growing interest into investigating market mechanisms that harness space-time flexiblity from DaCes. Recent work by Wierman and co-workers has identified various forms of DaCe flexibility that can be of practical use such as load shifting, load shedding, and geographical load balancing \citep{wierman_liu_liu_mohsenian-rad_2017}. The work in  \cite{ghatikar_ganti_matson_piette_2012} reviews how operations of DaCes can provide demand response services. The work in \cite{rao_liu_xie_liu_2010} analyzes how NOCs may be incentivized to exploit price differences across  electricity markets. The work in \cite{tran_tran_ren_han_huh_hong_2015} sets up a Stackelberg game to simulate optimal load-shifting strategies that DaCes might follow to real-time pricing signals. The work in \cite{liu_wierman_chen_razon_chen_2013} uses a stochastic optimization framework to study how DCs can predict coincidental peak-pricing and avoid high cost caused by peak hours using temporal workload shifting.  Research has also focused on harnessing DaCe flexibility for the specific purpose of exploiting the availability of renewable power (e.g., as a way to decarbonize operations and mitigate variability).  For example, \cite{kim_yang_zavala_chien_2017} shows that optimal placement and participation of DaCes in power grid markets (as dispatchable loads) leads to important reductions in power spillage and cost and to a better utilization of wind generation (thus enabling higher adoption levels). 

Modeling DaCe flexibility is challenging due to the complicated nature of their workloads \citep{wierman_liu_liu_mohsenian-rad_2017}. Specifically, DaCes need to process complex mixtures of flexible and inflexible loads that vary significantly over time. Work by \cite{liu_liu_low_wierman_2014} has shown that load-shifting in DaCes can be seen as a form of large-scale storage but also note that incentivizing the provision of DaCe flexibility in power markets is difficult. It is also important to recognize that different DaCes possess different types and degrees of spatial flexibility;  for instance, some DaCes might only possess local flexibility, while others might be equipped with geographical flexibility. Similarly, some DaCes have different levels of temporal flexibility, which is dictated by the nature of their workloads (e.g., job duration). 

DaCes demand responses have been studied using various game-theoretic modeling frameworks. For instance, \cite{zhou_li_wu_2018} and \cite{sun_ren_wu_li_2016} study how to incentivize spatial and temporal flexibility participation from the DaCe side using online auctioning. A Nash bargaining framework has been employed to study the design incentive mechanisms between data centers and load-serving entities \citep{cao_zhang_poor_2018} and between data centers and tenants in order to encourage the utilization of flexibility \citep{guo_li_pan_2018}. These studies have shown that the amount of flexibility that data centers are willing to offer is, in fact, highly dependent on electricity prices.  However, the decision-making process of DaCes is not properly taken into account in existing electricity market designs; as such, electricity prices are limited in their ability to incentivize the provision of flexibility. 

Power grids in the U.S. are operated using a coordinated market design where an independent system operator (ISO) collects bid information from power consumers and suppliers and uses this information to solve a market clearing problem (an optimization problem). The solution of this problem seeks to determine power allocations and locational marginal prices (LMPs) that maximize the social surplus (the collective profit of all players) subject to myriad constraints of the underlying physical assets (e.g., transmission, capacity). Clearing mechanisms in current use are largely based on the pioneering works \green{of Schweppe \cite{schweppe2013spot} and} of Hogan \citep{hogan_1992,hogan_read_ring_1996}, \green{the latter of }which establish revenue adequacy for transmission congestion payments and uses duality theory to determine proper remuneration mechanisms to participants using clearing prices. 

Various market clearing formulations have been proposed to capture different characteristics of specific types of assets. For instance, the work by \cite{carrion_arroyo_2006} proposes a clearing problem to capture time-dependent limitations (start-up costs and ramping constraints) in thermal units. The formulation proposed in \cite{bouffard_galiana_conejo_2005} aims to incorporate generation uncertainty directly in the clearing procedure. Along the same lines, the works in \cite{pritchard_zakeri_philpott_2010,zavala_kim_anitescu_birge_2017} propose stochastic clearing formulations to analyze whether LMPs properly reflect uncertainty of power generation, while maintaining key economic properties in the face of such uncertainty (e.g., revenue adequacy and cost recovery). The work in  \cite{gribik_chatterjee_navid_zhang_2011} proposed a pricing methodology that captures the cost of using an off-line resource to meet operational constraints. 
\green{Several works have explored extending the standard price-quantity bid format non-conventional market participants that are spatially and/or temporally flexible, including energy storage systems \citep{de2019implications} and prosumers \citep{ottesen2016prosumer}.
The work by \cite{liu2015extending} extends bid formats for adjustable, shiftable, and arbitrage loads. Recently, the concept of price-region bid format is introduced that captures general flexibility providers with linearly-constrained feasible region and convex piecewise linear cost \citep{bobo2021price}.}

\cite{virtual_link_pscc} have recently proposed a market clearing formulation that captures space-time load-shifting flexibility of DaCes. This flexibility is captured in the market clearing process using a load disaggregation procedure that can be represented using {\em virtual links}.  Virtual links are non-physical pathways that can be used by the ISO to shift power loads in space (i.e., by sending a computing load to another geographical location) and time (i.e., by delaying a computing load). This paradigm is compatible with existing market clearing procedures and reveals that virtual links form an additional infrastructure layer (similar to that overseen by NOCs) that complements the transmission network. {This paradigm also captures general spatially and temporally flexible loads that can be offered from other players such as distributed manufacturing facilities and storage systems.} {\em In this work}, we provide an in-depth theoretical analysis of market clearing formulations with virtual links. This reveals mechanisms under which DaCes should be remunerated for their flexibility and reveals information that DaCes should share with the ISO in the bidding process. This also shows that flexible consumers act as prosumers that simultaneously pay for requested load and that are paid for their flexibility service (there is an incentive to offer flexibility in order to decrease total cost). Our analysis also shows that virtual links provide a convenient mathematical construct to establish fundamental market properties; specifically, we show that a market that harnesses space-time load-shifting flexibility increases the social surplus, achieves revenue adequacy,  achieves cost recovery, and provides a mechanism to mitigate space-time volatility of LMPs. The framework is applied to case studies to illustrate the theoretical results and practical impact. 

The paper is structured as follows; Section \ref{sec:formulation} presents basic market clearing formulations that are used to introduce notation and concepts. Specifically, we consider a formulation that uses load disaggregation to capture DaCe flexibility and provide an equivalent representation that uses virtual links. Section \ref{sec:properties} presents a general market formulation that captures space-time virtual links along with its properties. Section \ref{sec:case_study} presents cases studies to illustrate the developments. Section \ref{sec:conclusion} closes with remarks on future directions.

\section{Basic Market Formulations} \label{sec:formulation}

The market clearing formulations under study incorporate several new elements that are not standard in the power systems literature. As such, we introduce the reader to these new concepts by exploring a family of formulations of increasing complexity. This will also allows us to introduce basic terminology, notation, and to highlight key concepts that motivate our work.  The market setting studied is for an {\em energy-only} setting, similar to those studied in recent market design work by \cite{pritchard_zakeri_philpott_2010}, \cite{kazempour2018stochastic}, \cite{zakeri2019pricing}, and \cite{zavala_kim_anitescu_birge_2017}.  

\subsection{Basic Notation and Terminology}

We begin our discussion by introducing basic notation for a static network (there is no time associated with it). The market considers a set of suppliers (owners of power plants) $\mathcal{S}$ and consumers (owners of DaCes) $\mathcal{D}$ connected to a transmission network comprised of geographical nodes $\mathcal{N}$ and transmission lines $\mathcal{L}$ (owned by transmission service providers).

Each supplier $i \in \mathcal{S}$ is connected to the power grid at node $n(i)\in \mathcal{N}$. The supplier bids into the market by offering power at bid price $\alpha_i^p \in \mathbb{R}_+$ and offers available capacity $\bar{p}_i \in [0, \infty)$. We define $\mathcal{S}_n := \{i \in \mathcal{S} \, | \, n(i) = n\} \subseteq \mathcal{S}$ (set of suppliers connected to node $n$). The cleared allocation for supplier $i\in \mathcal{S}$ (load injected) are denoted as $p_i$ and must satisfy $p_i \in [0,\bar{p}_i]$.  We use $p$ to denote the collection of all cleared allocations. 

A consumer $j \in \mathcal{D}$ bids into the market by requesting power at bid price $\alpha_j^d \in \mathbb{R}_+$ and requests a maximum capacity $\bar{d}_j \in [0, \infty)$. We define $\mathcal{D}_n := \{j \in \mathcal{D} \, | \, n(j) = n\} \subseteq \mathcal{D}$ (set of consumers connected to node $n$). For simplicity, we assume that there is only one consumer at a given node ($\mathcal{D}_n$ are singletons). The cleared allocation for consumer $j\in \mathcal{D}$ (load withdrawn) is denoted as $d_j$ and must satisfy $d_j \in [0,\bar{d}_j]$.  We use $d$ to denote the collection of all cleared allocations. A standard ({\em inflexible}) consumer requests that the cleared load $d_j$ is delivered at a single node $n(j)\in \mathcal{N}$. A {\em flexible} consumer (a DaCe owner), on the other hand, offers the possibility that the cleared load $d_j$ is delivered at a set of possible nodes $\mathcal{N}_d\subseteq{N}$; in other words, the cleared load $d_j$ can be disaggregated and individual portions are delivered at different nodes. We will see that this load disaggregation scheme can be seen as a spatial, load-shifting mechanism that can be modeled using {\em virtual links}. For simplicity, we will not make a distinction between DaCe owners and inflexible consumers. This is because an inflexible consumer can be modeled as a flexible consumer with $\mathcal{N}_d=\{n(j)\}$ (it offers one node option for the load to be delivered).  

Each transmission owner has a line $l \in \mathcal{L}$ defined by its sending node $\mathrm{snd}(l) \in \mathcal{N}$ and receiving node $\mathrm{rec}(l) \in \mathcal{N}$. The definitions of $\mathrm{snd}(l)$ and $\mathrm{rec}(l)$ are interchangeable because power can flow in either direction. The sending and receiving nodes of a link are also known as its supporting nodes. For each node $n \in \mathcal{N}$, we define its set of receiving lines $\mathcal{L}_n^{\textrm{rec}} := \{l \in \mathcal{L} \, | \, n=\textrm{rec}(l)\} \subseteq \mathcal{L}$ and its set of sending lines $\mathcal{L}_n^{\textrm{snd}} := \{l \in \mathcal{L} \, | \, n=\textrm{snd}(l)\}\subseteq \mathcal{L}$. Each line offers a bid price $\alpha^f_l \in \mathcal{R}_+$ and capacity $\bar{f}_l \in [0, \infty)$. {Note that while in common market cleraing practice, transmission line costs are not captured (i.e., $\alpha^f_l = 0$ for each line $l$), we add this generality in order to demonstrate the similarity between transmission network and flexibility network later. } Each cleared flow $f_l$ must satisfy the bounds $f_l \in [-\bar{f}_l, \bar{f}_l]$ and the collection $f$ must obey the direct-current (DC) power flow equations:
\begin{equation}
\label{eq:dc_flow}
f_l = B_l(\theta_{\textrm{snd}(l)} - \theta_{\textrm{rec}(l)}),
\end{equation}
where $B_l \in \mathbb{R}_+$ is the line susceptance and $\theta_n \in \mathbb{R}$ is the phase angle at node $n \in \mathcal{N}$. The DC power flow model is a linear model and requires small phase angle differences across transmission lines $\theta_{\textrm{snd}(l)} - \theta_{\textrm{rec}(l)} \in [-\Delta\bar{\theta'}_l,\Delta\bar{\theta'}_l]$. The limits on phase angle differences and the capacity constraints for flows can be captured as:
\begin{equation}
\label{eq:angle_diff_cap}
-\Delta\bar{\theta}_l \leq \theta_{\textrm{snd}(l)} - \theta_{\textrm{rec}(l)} \leq \Delta\bar{\theta}_l
\end{equation}
where $\Delta \bar{\theta}_l := \min\{\bar{f}_l/B_l, \Delta \bar{\theta'}_l \}$. 

We use $\pi_n\in \mathbb{R}_+$ to represent the cleared price at node $n \in \mathcal{N}$. The collection of cleared prices is denoted as $\pi$; these are also known as nodal prices or LMPs and are used to charge/remunerate market players. We observe that, in a typical market, suppliers and transmission owners {\em offer a service} to the grid, while inflexible consumers {\em request a service} from the grid. This distinction is important because we will see that flexible consumers (DaCes) act as {\em prosumers} that simultaneously request a service (request load) and offer a service (flexibility for load to be delivered at different locations); as such, a well-designed market should properly remunerate the provision of flexibility by DaCes. 

\subsection{Basic Formulation with Inflexible Consumers}

We begin our discussion by studying a clearing formulation with inflexible consumers:
\begin{subequations}
\label{opt:p1}
\begin{align}
& \underset{d,p,f,\theta}{\text{max}} && \sum_{j\in\mathcal{D}} \alpha_{j}^d d_{j} - \sum_{i \in \mathcal{S}} \alpha_{i}^p p_{i} - \sum_{l\in \mathcal{L}}\alpha_l^f|f_l| \label{opt:p1_obj} \\ 
& {\text{s.t.}} && \sum_{l \in \mathcal{L}_n^\textrm{rec}} f_l  + \sum_{i \in \mathcal{S}_n} p_{i} =  \sum_{l \in \mathcal{L}_n^\textrm{snd}} f_l + \sum_{j \in \mathcal{D}_n} d_{j}, \quad (\pi_n) \quad n \in \mathcal{N} \label{opt:p1_balance} \\ 
&&& f_l = B_l(\theta_{\textrm{snd}(l)} - \theta_{\textrm{rec}(l)}), \quad l \in \mathcal{L} \label{opt:p1_dc} \\ 
&&& d \in \mathcal{C}_d \, , \, p \in \mathcal{C}_p \, , \, \theta \in \mathcal{C}_\theta. \label{opt:p1_caps}
\end{align}
\end{subequations}
Here, we define the feasible capacity sets $\mathcal{C}_d := \{d\,| \, d_j \in [0, \bar{d}_j]\,\, \forall \,\, j \in \mathcal{D}\}$, $\mathcal{C}_p := \{p\,| \, p_i \in [0, \bar{p}_i]\,\, \forall \,\, i \in \mathcal{S}\}$ and $\mathcal{C}_\theta := \{\theta \, | \, \theta_{\textrm{rec}(l)} - \theta_{\textrm{snd}(l)} \in [-\Delta\bar{\theta}_l,\Delta\bar{\theta}_l] \,\, \forall \,\, l \in \mathcal{L}\}$. The objective function (\ref{opt:p1_obj}) is known as the {\em social surplus} or {\em total welfare}, which captures the value of demand served (to be maximized) and the cost of supply and transmission cost services (to be minimized).  The transmission cost is typically not included in the market clearing literature; this cost is included here to highlight an important analogy between transmission costs and load-shifting costs for DaCes (to be discussed later). Specifically, we will see that load-shifting creates an alternative, non-physical infrastructure network that is analogous to the transmission network. Constraint (\ref{opt:p1_balance}) is the power balance constraint at each node $n$ (Kirchoff's current law). 

The solution of the market clearing problem gives the {\em primal} allocations $(p,d,f)$ and the {\em dual} allocations $\pi$. The dual allocations are the dual variables associated with the power balance constraints \eqref{opt:p1_balance}. We will see that these can be used locational marginal prices (LMPs) that clear the market. We use $(p,d,f,\pi)$ to denote the primal-dual allocation obtained from the solution of the clearing formulation.  

The social surplus \eqref{opt:p1_obj} is a non-smooth function because of the presence of absolute value terms. As is standard practice \cite{intro_to_linear_opt}, this can be reformulated as a standard linear program by decomposing each line $l$ into directed edges $l^+ = (\textrm{snd}(l), \textrm{rec}(l))$, $l^- = (\textrm{rec}(l), \textrm{snd}(l))$ and replace the terms in \eqref{opt:p1} as $f_l \leftarrow f_{l^+} - f_{l^-}$, $|f_l| \leftarrow f_{l^+} + f_{l^-}$, and $f_{l^+} \geq 0, f_{l^-} \geq 0$.  We use $\mathcal{K}$ to represent the set of directed edges that results from this decomposition. Each edge $k \in \mathcal{K}$ resulted from line $l(k) \in \mathcal{L}$ inherits the susceptance $B_k := B_{l(k)}$, bid price $\alpha_k^f := \alpha_{l(k)}^f$, and capacity $\Delta \bar{\theta}_k := \Delta \bar{\theta}_{l(k)}$. 
\textcolor{black}{We define the profit term for each directed edge in a similar way:
\begin{equation}
	\phi^f_k(\pi_{\textrm{rec}(k)}, \pi_{\textrm{snd}(k)}, f_k) := (\pi_{\textrm{rec}(k)} - \pi_{\textrm{snd}(k)} - \alpha^f_k) f_k
\end{equation}
}
Using these definitions, the transmission cost can be expressed as:
\begin{equation}
\sum_{l\in \mathcal{L}}\alpha_l^f|f_l| = \sum_{l\in \mathcal{L}}\alpha_l^f(f_{l^+} + f_{l^-}) = \sum_{k\in\mathcal{K}}\alpha_k^f f_k, 
\end{equation}
and the net flows entering a node $n$ can be expressed as:
\begin{equation}
\begin{split}
\sum_{l\in\mathcal{L}^\textrm{rec}_n} f_l - \sum_{l\in\mathcal{L}^\textrm{snd}_n} f_l & = \sum_{l\in\mathcal{L}^\textrm{rec}_n} (f_{l^+} - f_{l^-}) - \sum_{l\in\mathcal{L}^\textrm{snd}_n} (f_{l^+} - f_{l^-}) \\ 
& = \left( \sum_{l\in\mathcal{L}^\textrm{rec}_n} f_{l^+} + \sum_{l\in\mathcal{L}^\textrm{snd}_n} f_{l^-} \right) - \left( \sum_{l\in\mathcal{L}^\textrm{rec}_n} f_{l^-} + \sum_{l\in\mathcal{L}^\textrm{snd}_n} f_{l^+} \right) \\
& = \sum_{k\in\mathcal{K}^\textrm{rec}_n} f_k - \sum_{l\in\mathcal{K}^\textrm{snd}_n} f_k. 
\end{split}
\end{equation}
This leads to the (equivalent) clearing problem:
\begin{subequations}
\label{opt:lp}
\begin{align}
& \underset{d,p,f,\theta}{\text{min}} && \sum_{i \in \mathcal{S}} \alpha_{i}^p p_{i} + \sum_{k\in \mathcal{K}}\alpha_k^ff_k - \sum_{j\in\mathcal{D}} \alpha_{j}^d d_{j}\label{opt:lp_obj} \\ 
& {\text{s.t.}} && \sum_{k \in \mathcal{K}_n^\textrm{rec}} f_k  + \sum_{i \in \mathcal{S}_n} p_{i} =  \sum_{k \in \mathcal{K}_n^\textrm{snd}} f_k + \sum_{j \in \mathcal{D}_n} d_{j}, \quad (\pi_n) \quad n \in \mathcal{N} \label{opt:lp_balance} \\ 
&&& f_{l^+} - f_{l^-} = B_l(\theta_{\textrm{snd}(l)} - \theta_{\textrm{rec}(l)}), \quad l \in \mathcal{L} \label{opt:lp_dc}\\ 
&&& d \in \mathcal{C}_d \, , \, p \in \mathcal{C}_p \, , \, \theta \in \mathcal{C}_\theta \label{opt:lp_caps}
\end{align}
\end{subequations}
In this formulation, we minimize the negative surplus (as opposed to maximize the surplus); this equivalent representation will facilitate the analysis. 

{A well-designed market clearing formulation must satisfy the following economic properties: 
\begin{itemize}
	\item {\em Competitive Equilibrium:} The clearing formulation must deliver allocations and prices that represent a competitive equilibrium. Specifically, the market must deliver allocations that balance supply and demand and that maximize the collective profit for all players. This property also ensures that the ISO does not interfere with the competitive nature of the market players. 
	\item {\em Revenue Adequacy:} The clearing formulation delivers allocations and prices such that the total amount of money paid by service requesters (consumers) covers the total amount paid to all service providers (suppliers and transmission). This also ensures that the ISO does not have financial gain. 
	\item {\em Cost Recovery:} The clearing formulation delivers allocations and prices such that no cleared player incurs a financial loss (it recovers its operating cost). 
\end{itemize}
Although not necessarily a fundamental property, it is often desired that prices delivered by the market are consistent with bid prices provided by market players (e.g., prices are bounded by bid prices provided by players). We will see that this property is intimately related to cost recovery; specifically, for a service provider to not incur a financial loss, its cleared price must be higher than its bid price (its marginal cost); for a consumer, the cleared price must be lower than its bid price. We thus see that prices must satisfy some inherent bounding properties in order for cost recovery to occur. 

\begin{figure}[!htb]
\centering
\includegraphics[width=0.8\textwidth]{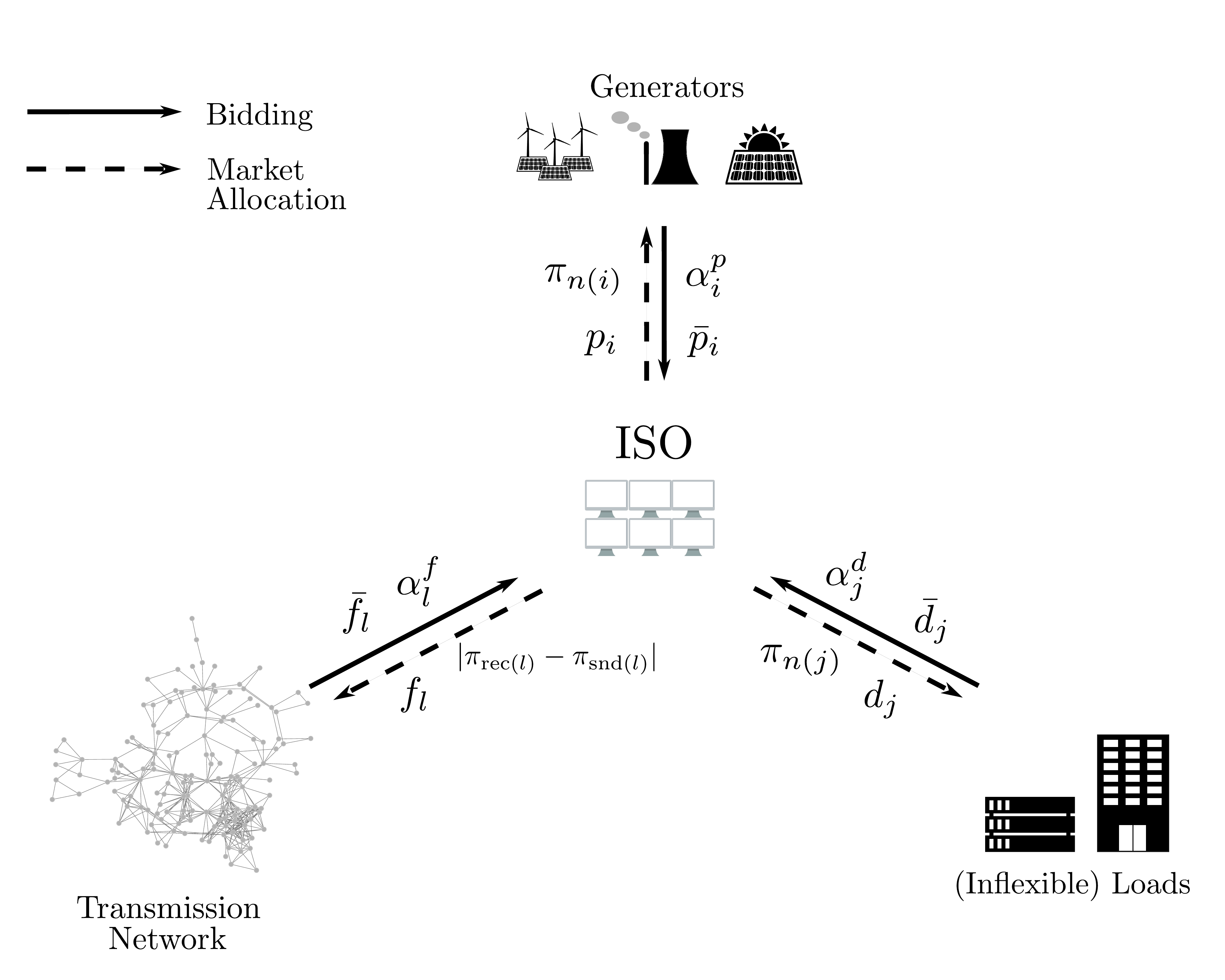}
\caption{\footnotesize Standard market clearing mechanism with inflexible loads.}
\label{fig:inflexible}
\end{figure}

We now show that the market clearing formulation satisfies the stated properties; our discussion here will be informal and is intended to introduce the general logic behind the analysis of clearing formulations (we will follow a similar logic in studying more complex formulations). 

We first need to define the  {\em mechanism} that will be used to charge/remunerate players and we then need to verify that such mechanism is compatible with the clearing formulation. We consider that each supplier $i$ is remunerated with price $\pi_{n(i)}$ for each unit of power cleared (injected) and each consumer $j$ pays $\pi_{n(j)}$ for each unit of power cleared (withdrawn). Each transmission provider $l$ is remunerated using the unit price $|\pi_{\textrm{rec}(l)} - \pi_{\textrm{snd}(l)}|$, which is the price difference between the supporting nodes. The profit functions for supplier $i \in \mathcal{S}$, consumer $j \in \mathcal{D}$, and transmission provider $l \in \mathcal{L}$ are thus:
\begin{subequations}
\begin{gather}
	\phi^p_i(\pi_{n(i)}, p_i) := (\pi_{n(i)} - \alpha^p_i) p_i \\
	\phi^d_j(\pi_{n(j)}, d_j) := (\alpha^d_j - \pi_{n(j)}) d_j \\
	\phi^f_l(\pi_{\textrm{rec}(l)}, \pi_{\textrm{snd}(l)}, f_l) := (|\pi_{\textrm{rec}(l)} - \pi_{\textrm{snd}(l)}| - \alpha^f_l) |f_l|
\end{gather}
\end{subequations}
When convenient, we use the short-hand notation $\phi^p_i,\phi^d_j$, and $\phi^f_l$.} We proceed by defining the (partial) Lagrange function of the clearing formulation \eqref{opt:lp}:
\begin{equation}
\label{eq:lagrange_lp}
\begin{aligned}
& L(\pi,d,p,f) & = & \sum_{i \in \mathcal{S}} \alpha_{i}^p p_{i} + \sum_{k \in \mathcal{K}}\alpha_k^f f_k - \sum_{j \in \mathcal{D}}  \alpha^d_jd_j -\sum_{n \in \mathcal{N}}\pi_n\left(\sum_{k \in \mathcal{K}_n^\textrm{rec}} f_k  + \sum_{i \in \mathcal{S}_n} p_{i} - \sum_{k \in \mathcal{K}_n^\textrm{snd}} f_k - \sum_{j \in \mathcal{D}_n} d_{j}\right) \\
&& = & -\sum_{j \in \mathcal{D}} (\alpha^d_j - \pi_{n(j)})d_j - \sum_{i \in \mathcal{S}}(\pi_{n(i)} - \alpha_i^p)p_i - \sum_{k \in \mathcal{K}}(\pi_{\textrm{rec}(k)} - \pi_{\textrm{snd}(k)} - \alpha^f_k) f_k \\
&& = & -\sum_{j \in \mathcal{D}} \phi^d_j - \sum_{i \in \mathcal{S}}\phi^p_i - \sum_{k \in \mathcal{K}}\phi^f_k,
\end{aligned}
\end{equation}
and the Lagrange dual function:
\begin{equation}
\label{eq:h}
\mathcal{D}(\pi) := \min_{d\in\mathcal{C}_d, p\in\mathcal{C}_p, f\in \mathcal{F}} L(\pi,d,p,f).
\end{equation}
where $\mathcal{F} := \{f \, | \, \exists \, \theta \in \mathcal{C}_\theta \, \text{s.t.} \, f_{l^+} - f_{l^-} = B_l(\theta_{\textrm{snd}(l)} - \theta_{\textrm{rec}(l)}) \, \forall \, l \in \mathcal{L}\}$ denotes the set of flows that satisfies constraints \eqref{opt:lp_dc}. These definitions allow us to formulate the Lagrangian dual problem: 
\begin{equation}
\label{opt:max_h}
\max_\pi \, \mathcal{D}(\pi).
\end{equation}
Throughout our study we assume that strong duality holds for all proposed market clearing formulations. For the setting discussed here, this guarantees that an optimal solution of the clearing problem \eqref{opt:lp}) can also be found by solving the corresponding Lagrangian dual problem \eqref{opt:max_h}.

We now proceed to show that the solution of the clearing formulation constitutes a {\em competitive equilibrium}. By definition, the solution satisfies the nodal balance constraints \eqref{opt:lp_balance}. Furthermore, given a set of prices $\pi$, the Lagrange dual function \eqref{eq:h} can be decomposed into the independent optimization problems $\max_{d_j \in [0, \bar{d}_j]} \, \phi^d_j$ for each consumer $j$, $\max_{p_i \in [0, \bar{p}_i]} \, \phi^p_i$ for each supplier $i$, and $\max_{f \in \mathcal{F}} \sum_{k\in\mathcal{K}} \phi^f_k$ for the transmission providers. Thus, the clearing formulation finds prices that maximize the profit function for each player. From strong duality we have that a solution of the Lagrangian dual problem also solves the clearing problem and thus satisfies the power balance constraints (supply equals demand at each node). 

To establish {\em revenue adequacy}, we need to show that the total revenue collected from consumers matches the total revenue allocated to service providers:
\begin{equation}
\sum_{j\in\mathcal{D}} \pi_n(j) d_j = \sum_{i\in\mathcal{S}}\pi_{n(i)} p_i + \sum_{k\in\mathcal{K}} (\pi_{\textrm{rec}(k)} - \pi_{\textrm{snd}(k)})f_k.
\end{equation}
From the power balance constraints \eqref{opt:lp_balance} we have that:
\begin{align}
 \sum_{k \in \mathcal{K}_n^{\textrm{rec}}} f_{k} + \sum_{i \in \mathcal{S}_n} p_{i} - \sum_{k \in \mathcal{K}_n^{\textrm{snd}}} f_{k} - \sum_{j \in \mathcal{D}_n} d_{j} = 0.
\end{align}
This implies that, 
\begin{align}
\sum_{n \in \mathcal{N}} \pi_n \left( \sum_{k \in \mathcal{K}_n^{\textrm{rec}}} f_{k} + \sum_{i \in \mathcal{S}_n} p_{i} - \sum_{k \in \mathcal{K}_n^{\textrm{snd}}} f_{k} - \sum_{j \in \mathcal{D}_n} d_{j} \right) = 0.
\end{align}
This expression can be written in the following equivalent form:
\begin{align}
\sum_{j \in \mathcal{D}} \pi_{n(j)} d_{j} &= \sum_{i \in \mathcal{S}} \pi_{n(i)} p_{i} + \sum_{n \in \mathcal{N}} \pi_{n} \left( \sum_{k \in \mathcal{K}_n^{\textrm{rec}}} f_{k} - \sum_{k \in \mathcal{K}_n^{\textrm{snd}}} f_{k} \right) \nonumber\\
&= \sum_{i \in \mathcal{S}} \pi_{n(i)} p_{i} + \sum_{k \in \mathcal{K}} \left(\pi_{\textrm{rec}(k)} - \pi_{\textrm{snd}(k)}\right) f_{k}, 
\end{align}
which establishes revenue adequacy. 

We now establish {\em cost recovery}; in establishing a competitive equilibrium, we argue that the Lagrangian dual problem \eqref{eq:h} maximizes the profit function for each individual player. Furthermore, since $(p,d,f,\theta) = (0,0,0,0)$ is a feasible (trivial) solution we have that, at the optimal allocation, the profit function for each player must be non-negative and thus $\phi^p_i,\phi^d_j,\phi^f_l\geq 0$. 

To see how cost recovery leads to price-boundedness, define the set of cleared suppliers $\mathcal{S}^* := \{i \in \mathcal{S} \, | \, p_i > 0\}$ and the set of cleared consumers $\mathcal{D}^* := \{j \in \mathcal{D} \, | \, d_j > 0\}$. From the argument behind cost recovery we have $\phi^p_i(\pi_{n(i)}, p_i) = (\pi_{n(i)} - \alpha^p_i) p_i \geq 0$ and $\phi^d_j(\pi_{n(j)}, d_j) = (\alpha^d_j - \pi_{n(j)}) d_j\geq 0$. For $i \in \mathcal{S}^*$, we have $p_i > 0$ and thus $\pi_{n(i)} \geq \alpha^p_i$. Similarly, for $j\in \mathcal{D}^*$, we have $d_j>0$ and thus $\pi_{n(j)} \leq \alpha^d_j$. 

It is often observed that increasing transmission capacity of a line has the effect of reducing the price difference between the connected nodes; in other words, transmission capacity reduces the {\em spatial variability} of prices. In the limit when there is enough transmission capacity and the network is well-connected, power should be allowed to move freely in the network (there is no market friction) and all nodal prices should collapse to a single value. On the other hand, when there is not enough transmission capacity and/or the network is not well-connected, there will be large differences between nodal prices. Understanding the effect of transmission capacity on prices will become relevant for the formulations studied in this paper; specifically, we will see that virtual links form an alternative infrastructure network that can help overcome limitations of the transmission network. 

To gain some intuition into how transmission capacity reduces spatial volatility, here we present a simple analysis in the absence of DC constraints \eqref{opt:lp_dc}; in such a case, the clearing formulation reduces to:  
\begin{subequations}
\label{opt:lp_nodc}
\begin{align}
& \underset{d,p,f}{\text{min}} && \sum_{i \in \mathcal{S}} \alpha_{i}^p p_{i} + \sum_{k\in \mathcal{K}}\alpha_k^ff_k - \sum_{j\in\mathcal{D}} \alpha_{j}^d d_{j} \label{opt:lp_nodc_obj} \\ 
& {\text{s.t.}} && \sum_{k \in \mathcal{K}_n^\textrm{rec}} f_k  + \sum_{i \in \mathcal{S}_n} p_{i} =  \sum_{k \in \mathcal{K}_n^\textrm{snd}} f_k + \sum_{j \in \mathcal{D}_n} d_{j}, \quad (\pi_n) \quad n \in \mathcal{N} \label{opt:lp_nodc_balance} \\ 
&&& d \in \mathcal{C}_d \, , \, p \in \mathcal{C}_p \, , \, f \in \mathcal{C}_f \label{opt:lp_nodc_caps}
\end{align}
\end{subequations}
where $\mathcal{C}_f = \{f|f_k \in [0, \bar{f}_k]\}$ and $\bar{f}_k = B_k\bar{\theta}_k$.  Consider now a couple of  instances for the clearing formulation \eqref{opt:lp_nodc} with line capacities $\bar{f}_k$ and $\bar{f}'_k$ satisfying $\bar{f}_k < \bar{f}'_k$ for some $k \in \mathcal{K}$. Let $\pi^*, \pi'^*$ be the optimal prices for the corresponding instances; we would like to show that the prices satisfy:
\begin{equation}
\pi'^*_{\textrm{rec}(k)} - \pi'^*_{\textrm{snd}(k)}\leq \pi^*_{\textrm{rec}(k)} - \pi^*_{\textrm{snd}(k)}.
\end{equation}
The Lagrange function of \eqref{opt:lp_nodc} has the same form as that of \eqref{eq:lagrange_lp}. The Lagrange dual function is:
\begin{equation}
\label{eq:h_nodc}
\mathcal{D}(\pi) = \min_{d\in\mathcal{C}_d, p\in\mathcal{C}_p, f\in\mathcal{C}_f} L(\pi,d,p,f).
\end{equation}
For fixed prices $\pi$, the Lagrange function is separable and be decomposed into the individual profit-maximization subproblems:
\begin{subequations}
\label{opt:load_max1}
\begin{alignat}{3}
    \max_{d} \quad & (\alpha^d_j - \pi_{n(j)} )d_j \\
    \text{s.t.} \quad & 0\leq d_n \leq \Bar{d}_j
\end{alignat}
\end{subequations}
\begin{subequations}
\label{opt:gen_max1}
\begin{alignat}{3}
    \max_{p_i} \quad &  (\pi_{n(j)} - \alpha^p_i)p_i \\
    \text{s.t.} \quad & 0 \leq p_i \leq \Bar{p}_i
\end{alignat}
\end{subequations}
\begin{align}\label{opt:trans_max1}
    \max_{f} \quad & \sum_{k \in \mathcal{K}}(\pi_{\textrm{rec}(k)} - \pi_{\textrm{snd}(k)} - \alpha^f_k) f_k.
\end{align}
These subproblems have explicit solutions and this allows us to write the Lagrangian dual function as:
\begin{equation}
\mathcal{D}(\pi) = -\sum_{j \in \mathcal{D}} |\alpha^d_j - \pi_{n(j)}|_+\bar{d}_j - \sum_{i \in \mathcal{S}}|\pi_{n(i)} - \alpha_i^p|_+\bar{p}_i - \sum_{k \in \mathcal{K}}|\pi_{\textrm{rec}(k)} - \pi_{\textrm{snd}(k)} - \alpha^f_k|_+ \bar{f}_k
\end{equation}
where $|\cdot|_+ = \max\{\cdot, 0\}$. The Lagrangian dual problem is: 
\begin{equation}
\max_\pi \, \mathcal{D}(\pi).
\end{equation}
Defining $\mathcal{D}'(\pi)$ as the Lagrangian dual function for the problem with expanded capacity $\bar{f}'_k$; we have that $h'(\pi) \leq \mathcal{D}(\pi)$ for any $\pi$ due to the larger feasible region of the inner problem.  Now let $( d^*, p^*, f^*,\pi^*)$ be an optimal solution for $\max_{\pi} \mathcal{D}(\pi)$, and define $\Delta_k(\pi) := \pi_{\textrm{rec}(v)} - \pi_{\textrm{snd}(v)} - \alpha^f_k$.  We consider a couple of cases; for the first case, assume $\Delta_k(\pi^*) \leq 0$, we thus have $\mathcal{D}'(\pi^*) = \mathcal{D}(\pi^*)$ and thus the optimal prices remain the same; for the second case, assume $\Delta_k(\pi^*) > 0$ and let $\pi$ be arbitrary prices such that $\Delta_k(\pi) > \Delta_k(\pi^*) > 0$. We now observe that $\pi$ cannot be optimal; from the setup above we have that:
\begin{subequations}
\begin{gather}
\mathcal{D}(\pi) - \mathcal{D}'(\pi) = \Delta_k(\pi) (\bar{f}'_k - \bar{f}_k) \\
\mathcal{D}(\pi^*) - \mathcal{D}'(\pi^*) = \Delta_k(\pi^*) (\bar{f}'_k - \bar{f}_k),
\end{gather}
\end{subequations}
this implies that:
\begin{equation}
\begin{aligned}
\mathcal{D}'(\pi) - \mathcal{D}'(\pi^*) = \mathcal{D}(\pi) - \mathcal{D}(\pi^*)+(\bar{f}'_k - \bar{f}_k) (\Delta_k(\pi^*) - \Delta_k(\pi)) < 0,
\end{aligned}
\end{equation}
which holds because $\Delta_k(\pi^*) - \Delta_k(\pi) < 0$ and $\mathcal{D}(\pi) \leq \mathcal{D}(\pi^*)$ (by optimality of $\pi^*$). 

This simple analysis provides some intuition into how the addition of transmission flexibility can help mitigate nodal price differences. We will see later that a similar behavior can be obtained by incorporating spatial load-shifting flexibility provided by DaCes.  Moreover, we will see that temporal load-shifting flexibility can be used to mitigate temporal price differences. 


\subsection{Basic Formulation with Flexible Consumers}

We now expand the previous clearing formulation by incorporating flexible consumers; specifically, we consider flexible consumers that offer load-shifting flexibility by exploiting the availability of multiple DaCes placed at different nodes.  For simplicity in the presentation, here we consider a single flexible consumer. Suppose that this consumer submits a bid for requested load $\Bar{d}$ with bid price $\alpha^d$; the requested load can be served/delivered at a set of nodes $\mathcal{N}_d \subseteq \mathcal{N}$. Each node $n \in \mathcal{N}_d$ receives a partial load $d_n\geq 0$; the total load served to the consumer is $\sum_{n \in \mathcal{N}_d} d_n$ and satisfies $\sum_{n \in \mathcal{N}_d} d_n \leq \Bar{d}$ (total load served cannot exceed the requested load).  The clearing problem is: 
\begin{subequations}
\label{opt:disagg}
\begin{align}
& \underset{d,p,f,\theta}{\text{min}} && \sum_{i \in \mathcal{S}} \alpha_{i}^p p_{i} + \sum_{k \in \mathcal{K}}\alpha_k^f f_k - \alpha^d \sum_{n \in \mathcal{N}_d} d_n \label{opt:disagg_obj} \\
& {\text{s.t.}} && \sum_{k \in \mathcal{K}_n^{\textrm{rec}}} f_k + \sum_{i \in \mathcal{S}_n} p_{i} = \sum_{k \in \mathcal{K}_n^{\textrm{snd}}} f_k + d_n \quad (\pi_n),  \quad n \in \mathcal{N}_d \label{opt:disagg_balance_d} \\ 
&&& \sum_{k \in \mathcal{K}_n^{\textrm{rec}}} f_k + \sum_{i \in \mathcal{S}_n} p_{i} = \sum_{k \in \mathcal{K}_n^{\textrm{snd}}} f_k \quad (\pi_n),  \quad n \in \mathcal{N}\backslash\mathcal{N}_d \label{opt:disagg_balance_nod} \\ 
&&& 0 \leq \sum_{n \in \mathcal{N}_d} d_n \leq \Bar{d} \label{opt:disagg_total_d} \\
&&& d_n \geq 0,\quad n \in \mathcal{N}_d \label{opt:disagg_dn} \\
&&& f_{l^+} - f_{l^-} = B_l(\theta_{\textrm{snd}(l)} - \theta_{\textrm{rec}(l)}), \quad l \in \mathcal{L} \label{opt:disagg_kvl} \\
&&& p \in \mathcal{C}_p \, , \, \theta \in \mathcal{C}_{\theta} \label{opt:disagg_caps}
\end{align}
\end{subequations}
From the structure of the surplus function we see that the formulation aims to maximize the {\em total load} delivered to the flexible consumer. From the power balances, we see that nodes offered by the flexible consumer ($\mathcal{N}_d$) can receive load, while those not offered ($\mathcal{N}\setminus \mathcal{N}_d$) cannot. The profit function for the flexible consumer is defined as $\phi^d(\pi,d) := \sum_{n\in \mathcal{N}_d} (\alpha^d - \pi_{n} )d_n$. This indicates that the consumer is charged for power based on the prices of all the nodes offered. This remuneration mechanism is fundamentally different from that of an inflexible consumer (which is charged based on the price at a single node). As such, when a consumer offers load to be delivered at multiple nodes, it is expected that the ISO can exploit this flexibility to maximize the consumer profit (if this is not the case, there is no incentive for the consumer to offer flexibility).

\textcolor{black}{We now establish the fundamental market properties for \eqref{opt:disagg} to highlight differences with the previous market setting.} The partial Lagrange function is:
\begin{equation}
\label{eq:disagg_lagrange}
\begin{aligned}
L(\pi,d,p,f) & = && \sum_{i \in \mathcal{S}} \alpha_{i}^p p_{i} + \sum_{k \in \mathcal{K}}\alpha_k^f f_k - \alpha^d \sum_{n \in \mathcal{N}_d} d_n -\sum_{n \in \mathcal{N}\backslash \mathcal{N}_d}\pi_n\left(\sum_{k \in \mathcal{K}_n^{\textrm{rec}}} f_k - \sum_{i \in \mathcal{S}_n} p_{i} - \sum_{k \in \mathcal{K}_n^{\textrm{snd}}} f_k\right) \\
&&& \qquad -\sum_{n \in \mathcal{N}_d}\pi_n\left(\sum_{k \in \mathcal{K}_n^{\textrm{rec}}} f_k - \sum_{i \in \mathcal{S}_n} p_{i} - \sum_{k \in \mathcal{K}_n^{\textrm{snd}}} f_k - d_n\right) \\
& = && -\sum_{j \in \mathcal{N}_d} (\alpha^d - \pi_{n(j)})d_j - \sum_{i \in \mathcal{S}}(\pi_{n(i)} - \alpha_i^p)p_i - \sum_{k \in \mathcal{K}}(\pi_{\textrm{rec}(k)} - \pi_{\textrm{snd}(k)} - \alpha^f_k) f_k,
\end{aligned}
\end{equation}
and the Lagrangian dual problem is:
\begin{subequations}
\begin{alignat}{3}
    & \max_\pi \, \mathcal{D}(\pi) := \quad & \min_{d,p,f,\theta} \quad & L(\pi,d,p,f) \\
    && \text{s.t.} \quad & 0 \leq \sum_{j\in \mathcal{N}_d}d_j \leq \Bar{d} \\
    &&& d_n \geq 0, \quad n \in \mathcal{N}_d \\
    &&& f_{l^+} - f_{l^-} = B_l(\theta_{\textrm{snd}(l)} - \theta_{\textrm{rec}(l)}), \quad l \in \mathcal{L} \\
    &&& p \in \mathcal{C}_p \, , \, \theta \in \mathcal{C}_{\theta}
\end{alignat}
\end{subequations}
\textcolor{black}{
The Lagrange dual function $\mathcal{D}(\pi)$ can be decomposed to individual profit maximization problems:
\begin{subequations}
\label{opt:load_max}
\begin{alignat}{3}
    \max_{d} \quad & \sum_{j\in \mathcal{N}_d} (\alpha^d - \pi_{n(j)} )d_j \\
    \text{s.t.} \quad & 0 \leq \sum_{j\in \mathcal{N}_d}d_j \leq \Bar{d} \\
    & d_n \geq 0, \quad n \in \mathcal{N}_d
\end{alignat}
\end{subequations}
\begin{subequations}
\label{opt:gen_max}
\begin{alignat}{3}
    \max_{p_i} \quad &  (\pi_{n(j)} - \alpha^p_i)p_i \\
    \text{s.t.} \quad & 0 \leq p_i \leq \Bar{p}
\end{alignat}
\end{subequations}
\begin{subequations}
\label{opt:trans_max}
\begin{alignat}{3}
    \max_{f,\theta\in \mathcal{C}_{\theta}} \quad & \sum_{k \in \mathcal{K}}(\pi_{\textrm{rec}(k)} - \pi_{\textrm{snd}(k)} - \alpha^f_k) f_k \\
    \text{s.t.} \quad & f_{l^+} - f_{l^-} = B_l(\theta_{\textrm{snd}(l)} - \theta_{\textrm{rec}(l)}), \quad l \in \mathcal{L}
\end{alignat}
\end{subequations}
}
It is straightforward to show that the profit maximization problems for suppliers \eqref{opt:gen_max} and transmission providers \eqref{opt:trans_max} are the same as those of the base formulation with inflexible consumers (under DC constraints). The main difference that arises here lies in the profit maximization problem for the consumer \eqref{opt:load_max}; the structure of this problem confirms that the flexible consumer should be charged based on the nodal prices and the corresponding components of the disaggregated load; moreover, the total load served should not exceed the requested capacity.  

We now establish the properties for the market clearing formulation; the logic is similar as that followed before but we highlight some basic differences that arise from the provision of flexibility.  Market clearance is guaranteed by satisfaction of the power balance constraints \eqref{opt:disagg_balance_d}, \eqref{opt:disagg_balance_nod}. Furthermore, \eqref{opt:load_max}-\eqref{opt:trans_max} show that the formulation \eqref{opt:disagg} delivers an optimal price and allocation that maximize profit for the flexible consumer, each of the suppliers, and the transmission network.  From the balance constraints \eqref{opt:disagg_balance_d}, \eqref{opt:disagg_balance_nod} we have:
\begin{equation}
0 = \sum_{n \in \mathcal{N}} \pi_n \left( \sum_{k \in \mathcal{K}_n^{\textrm{rec}}} f_{k} + \sum_{i \in \mathcal{S}_n} p_{i} - \sum_{k \in \mathcal{K}_n^{\textrm{snd}}} f_{k} - \sum_{j \in \mathcal{D}_n} d_{j} \right)
\end{equation}
Basic manipulations reveal that this expression implies revenue adequacy. 

To establish cost recovery, we note again that $(p,d,f,\theta) = (0,0,0,0)$ is a feasible solution and thus the profit function of all players is non-negative. Non-negative profits implies that $\phi^p_i(\pi_{n(i)}, p_i) = (\pi_{n(i)} - \alpha^p_i) p_i \geq 0$ and $\phi^d(\pi,d) = \sum_{j\in \mathcal{N}_d} (\alpha^d - \pi_{n(j)} )d_j \geq 0$. If supplier $i$ satisfies $p_i > 0$, then $\pi_{n(i)} \geq \alpha^p_i$. To establish upper bounds,  suppose by contradiction that there exists $n \in \mathcal{N}_d$ such that $\alpha^d - \pi_n < 0$ and $d_n > 0$; then, $d$ is not optimal ($d$ does not attain the maximum profit for the flexible consumer); we can construct $d'$ by letting $d'_{n'} = d_{n'}$ for $n' \in \mathcal{D} \backslash \{j\}$ and $d'_n = 0$. We thus have that $d'$ satisfies all constraints and gives a higher profit; as such, $d_n > 0$ implies $\pi_n \leq \alpha^d$. We thus see that the {\em introduction of flexibility affects price boundedness}; specifically, for markets with inflexible loads, a nodal price can only be upper bounded by the bid price of the load connected to it. This is not the case for markets with flexible loads; specifically, the price for any node in $\mathcal{N}_d$ can be bounded by the load bid price $\alpha^d$. Therefore, the key insight here is that load-shifting flexibility provides a new mechanism for the ISO to control price behavior.

\begin{figure}[!htp]
\centering
\includegraphics[width=0.8\textwidth]{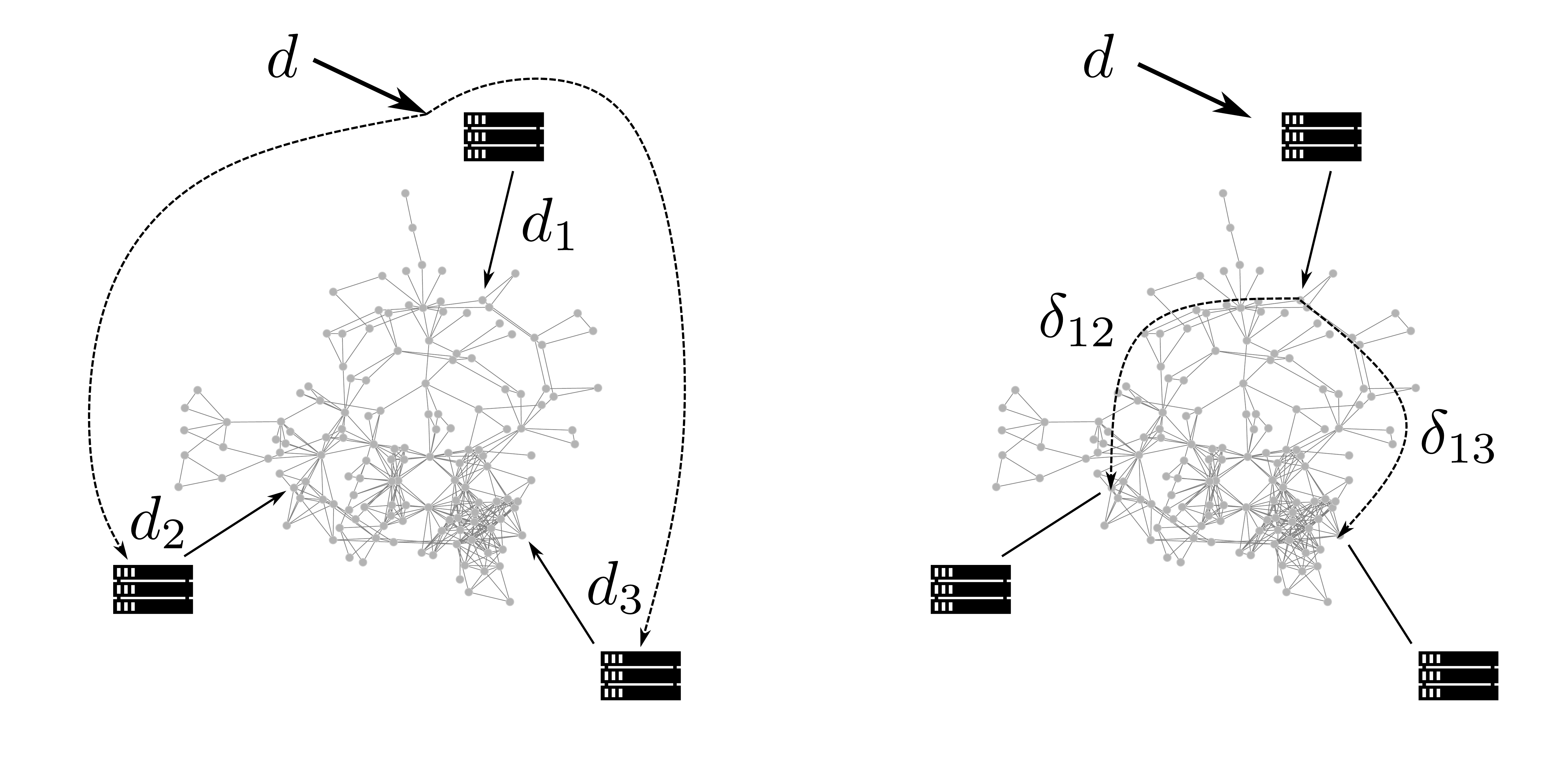}
\vspace{-0.1in}
\caption{\footnotesize Disaggregation model (left) and equivalent representation using virtual links  (right).}
\label{fig:disag}
\end{figure}

\subsection{Basic Formulation with Virtual Links}
\label{sec:basic_vl_market}
We have seen that there natural incentive for flexible consumers to offer alternative nodes to the ISO in order to access alternative nodal prices. However, the market clearing formulation previously explored does not provide intuition on how DaCe flexibility is remunerated.  To address this issue, we propose a mathematically-equivalent formulation that treats DaCes as prosumers that simultaneously request power and offer load-shifting flexibility. To do this, we introduce the notion of virtual links; specifically, as shown in Figure \ref{fig:disag}, load disaggregation can be seen as a non-physical transport (shift) of load from a reference node to a set of alternate nodes.  

Suppose the flexible consumer submits a bid for requested load $\Bar{d}$ at a {\em hub} (reference) node $n_h\in \mathcal{N}$ with bid price $\alpha^d$ and offers at set of alternate nodes $\mathcal{N}_d\subseteq \mathcal{N}$ such that $n_h\in \mathcal{N}_d$. We use $\delta_{n_h,n}\in \mathbb{R}_+$ to denote the amount of load that is shifted from the hub node $n_h$ to the alternate node $n \in \mathcal{N}_d$; we refer to the load-shifting pathway as a {\em virtual link}. This leads to the following clearing formulation:
\begin{subequations}
\label{opt:disagg_vl}
\begin{align}
& \underset{d,p,f,\theta,\delta}{\text{min}} && \sum_{i \in \mathcal{S}} \alpha_{i}^p p_{i} + \sum_{k \in \mathcal{K}}\alpha_k^f f_k - \alpha^d d \label{opt:disagg_vl_obj} \\
& {\text{s.t.}} && \sum_{k \in \mathcal{K}_n^{\textrm{rec}}} f_k + \sum_{i \in \mathcal{S}_n} p_{i} = \sum_{k \in \mathcal{K}_n^{\textrm{snd}}} f_k +
\begin{cases}
	d - \sum_{j \in \mathcal{N}_d} \delta_{n_h, j}, \quad n = n_h \\
	\delta_{n_h, n}, \quad n \in \mathcal{N}_d\backslash \{n_h\} \\
	0, \quad n \in \mathcal{N}\backslash \mathcal{N}_d
\end{cases}
\quad (\pi_n) \label{opt:disagg_vl_balance} \\ 
&&& 0 \leq d \leq \bar{d} \label{opt:disagg_vl_total_load_bounds} \\
&&& \delta_n \geq 0, \quad n \in \mathcal{N}_d \label{opt:disagg_vl_balance_vlcap} \\
&&& d - \sum_{n \in \mathcal{N}_d} \delta_{n_h,n} \geq 0 \label{opt:disagg_vl_balance_hub_cap} \\
&&& f_{l^+} - f_{l^-} = B_l(\theta_{\textrm{snd}(l)} - \theta_{\textrm{rec}(l)}), \quad l \in \mathcal{L} \label{opt:disagg_vl_kvl} \\
&&& p \in \mathcal{C}_p \, , \, \theta \in \mathcal{C}_{\theta} \label{opt:disagg_vl_caps}
\end{align}
\end{subequations}
It is not difficult to observe that formulations \eqref{opt:disagg} and \eqref{opt:disagg_vl} are equivalent. Specifically, there exists a bijection between $d_n$ in (\ref{opt:disagg}) and $(d,\delta_{n_h,n})$ in (\ref{opt:disagg_vl}) such that each pair satisfies:
\begin{gather*}
	d_{n_h} = d - \sum_{n \in \mathcal{N}_d}\delta_{n_h,n} \\
	d_n = \delta_{n_h,n}, \quad n \in \mathcal{N}_d\backslash \{n_h\}.
\end{gather*}
A feasible solution $(d,p,f,\theta,\delta)$ of (\ref{opt:disagg_vl}) implies existence of a feasible solution $(d,p,f,\theta)$ for \eqref{opt:disagg} with the same value of $(f,\theta,p)$ (and viceversa). In addition, both solutions attain the same optimal objective value. The partial Lagrange function for \eqref{opt:disagg_vl} is:
\begin{equation}
\label{eq:lagrange_vl_single}
\begin{aligned}
& L(d,p,\theta,\delta) & = & \sum_{i \in \mathcal{S}} \alpha_{i}^p p_{i} + \sum_{k \in \mathcal{K}}\alpha_k^f f_k - \alpha^d d - \pi_{n_h}\Big(\sum_{k \in \mathcal{K}_{n_h}^{\textrm{rec}}} f_k + \sum_{i \in \mathcal{S}_{n_h}} p_{i} - \sum_{k \in \mathcal{K}_{n_h}^{\textrm{snd}}} f_k - d + \sum_{n \in \mathcal{N}_d} \delta_{n_h,n}\Big) \\
&&& \quad - \sum_{n \in \mathcal{N}_d \setminus \{n_d\}}\pi_n \Big( \sum_{k \in \mathcal{K}_n^{\textrm{rec}}} f_k + \sum_{i \in \mathcal{S}_n} p_{i} - \sum_{k \in \mathcal{K}_n^{\textrm{snd}}} f_k - \delta_{n_h,n} \Big) \\
&&& \quad - \sum_{n \in \mathcal{N} \setminus \mathcal{N}_d \setminus \{n_d\}}\pi_n\Big(\sum_{k \in \mathcal{K}_n^{\textrm{rec}}} f_k + \sum_{i \in \mathcal{S}_n} p_{i} - \sum_{k \in \mathcal{K}_n^{\textrm{snd}}} f_k \Big) \\
&& = & (\pi_{n_h}-\alpha^d )d - \sum_{n \in \mathcal{N}_d} (\pi_{n_h} - \pi_n)\delta_{n_h,n} - \sum_{i \in \mathcal{S}}(\pi_i - \alpha_i^p)p_i - \sum_{k \in \mathcal{K}}(\pi_{\textrm{rec}(k)} - \pi_{\textrm{snd}(k)} - \alpha^f_k) f_k,
\end{aligned}
\end{equation}
and the Lagrangian dual problem is:
\begin{subequations}
\label{opt:lagrange_vl}
\begin{alignat}{3}
    & \max_\pi \quad & \min_{d,p,f,\theta} \quad & L(\pi,d,p,\theta) \\
    && \text{s.t.} \quad & 0 \leq d \leq \Bar{d} \\
    &&& \delta_{n_h,n} \geq 0, \quad n \in \mathcal{N}_d \\
    &&& d - \sum_{n \in \mathcal{N}_d} \delta_{n_h,n} \geq 0 \\
    &&& f_{l^+} - f_{l^-} = B_l(\theta_{\textrm{snd}(l)} - \theta_{\textrm{rec}(l)}), \quad l \in \mathcal{L} \\
    &&& p \in \mathcal{C}_p \, , \, \theta \in \mathcal{C}_{\theta}
\end{alignat}
\end{subequations}
The Lagrange function (\ref{eq:lagrange_vl_single}) reveals that virtual links are a {\em service} offered by the consumer. The market remunerates the consumer for the provision of this service via the profit 
$\sum_{n \in \mathcal{N}_d} (\pi_{n_h} - \pi_n)\delta_{n_h,n}$. This highlights that load-shifting is incentivized whenever there is a nodal price $\pi_n$ that is lower than the price at the hub node $\pi_{n_h}$. This is analogous to how transmission is remunerated (based on nodal price differences). The market also charges the consumer via the the profit $(\alpha^d - \pi_{n_h})d$; consequently, the flexible consumer acts as a prosumer and has total profit: 
\begin{equation}
(\alpha^d - \pi_{n_h})d + \sum_{n \in \mathcal{N}_d} (\pi_{n_h} - \pi_n)\delta_{n_h,n} = (\alpha^d - \pi_{n_h}) (d - \sum_{n\in\mathcal{N}_d}\delta_{n_h,n}) + \sum_{n\in\mathcal{N}_d}(\alpha^d - \pi_n) \delta_{n_h,n}
\end{equation}
This is the same profit function shown in \eqref{eq:disagg_lagrange}, where the profit is determined by the difference between the bid price and the price at the nodes where the loads are shifted to (this further reinforces the equivalence between the load disaggregation formulation and the formulation with virtual links).  The load disaggregation formulation shows the total remuneration for DaCes, while the formulation with virtual links reveals how the market remunerates the provision of load-shifting services.  We observe that all market and price properties established for the load disaggregation model hold for the virtual link model (since the models are equivalent); as such, these are not established again. 

\subsection{Basic Formulation with General Virtual Links (Spatial)}

We now generalize the concept of virtual links as a means to offer spatial load-shifting flexibility services; this will reveal strong connections between the non-physical network formed by virtual links and the physical transmission network. Specifically, we will see that virtual links form an additional infrastructure layer that is not restricted by DC power flow laws.  

We let $\mathcal{V}$ be the set of all virtual links; each virtual link $v \in \mathcal{V}$ has an associated bid price $\alpha^\delta_v \in \mathbb{R}_+$ and capacity $\bar{\delta}_v \in [0, \infty)$. The cleared load shifts (virtual flows) are defined as $\delta_v \in \mathbb{R}_+$ and are subject to capacity constraints $\delta_v \in [0,\bar{\delta}_v]$. We define $\mathcal{V}_n^\textrm{snd} := \{v\in\mathcal{V} \,|\, \textrm{snd}(v) = n\} \subseteq \mathcal{V}$, $\mathcal{V}_n^\textrm{rec} := \{v\in\mathcal{V} \,|\, \textrm{rec}(v) = n\} \subseteq \mathcal{V}$ to be the set of sending and receiving virtual links at node $n$. \textcolor{black}{Each flexible consumer $j \in \mathcal{D}$ is associated with a set of virtual links $\mathcal{V}_j\subseteq \mathcal{V}$. For each $v \in \mathcal{V}_j$, $\textrm{snd}(v) = n_h(j)$ since each consumer can only bid flexibility going from its hub node to other alternative nodes.} We note that the bid price of the virtual link can represent cost of shifting load (e.g., opportunity cost of migrating a computing load). This can also help capture the fact that shifts to certain nodes can be more expensive (e.g., due to distance or to capture preferred locations by the market players).  The shift cost is analogous to the service cost of power transmission. 

In a base setting with inflexible consumers, the load cleared (withdrawn) at a node $n$ is $\hat{d}_n = \sum_{j \in \mathcal{D}_n} d_j$. This does not hold if we consider flexible consumers, as a requested load at a given node might be withdrawn at another node. We thus have that the load withdrawn at node $n$ is:
\begin{align}
\hat{d}_n = \sum_{j \in \mathcal{D}_n} d_j + \sum_{v \in \mathcal{V}^{in}_n} \delta_v - \sum_{v \in \mathcal{V}^{out}_n} \delta_v.
\end{align}
As load shifting is introduced to the market clearing, the clearing process needs to ensure that a certain DaCe does not absorb a load that exceeds its available computing capacity. Moreover, the clearing process needs to ensure that a certain DaCe does not shift load that exceeds how much it actually possesses. This logic can be captured using the {\em computing capacity constraints}:
\begin{equation}
0 \leq \sum_{j \in \mathcal{D}_n}d_j + \sum_{v\in\mathcal{V}^{\textrm{rec}}_n}\delta_v - \sum_{v\in\mathcal{V}^{\textrm{snd}}_n} \delta_v \leq \bar{d}^{\max}_n, \quad n \in \mathcal{N}
\end{equation}
where $\bar{d}_n^{\textrm{max}}$ denotes the capacity of the DaCe located at $n$.

The market clearing problem with spatial virtual links is:
\begin{subequations}
\label{opt:space}
\begin{align}
& \underset{d,p,f,\theta,\delta}{\text{min}} && \sum_{i \in \mathcal{S}} \alpha_{i}^p p_{i} + \sum_{k\in \mathcal{K}}\alpha_k^f f_k + \sum_{v \in \mathcal{V}} \alpha^\delta_v \delta_v - \sum_{j\in\mathcal{D}} \alpha_{j}^d d_{j} \label{opt:space_obj} \\ 
& {\text{s.t.}} && \sum_{k \in \mathcal{K}_n^{\textrm{rec}}} f_k + \sum_{i \in \mathcal{S}_n} p_{i} + \sum_{v \in \mathcal{V}^{\textrm{snd}}_n} \delta_v = \sum_{k \in \mathcal{K}_n^{\textrm{snd}}} f_k + \sum_{j \in \mathcal{D}_n} d_{j} + \sum_{v \in \mathcal{V}^{\textrm{rec}}_n} \delta_v, \, (\pi_n) \, n \in \mathcal{N} \label{opt:space_balance} \\ 
&&& f_{l^+} - f_{l^-} = B_l(\theta_{\textrm{snd}(l)} - \theta_{\textrm{rec}(l)}), \quad l \in \mathcal{L} \label{opt:space_dtheta_kvl} \\
&&& 0 \leq \sum_{j \in \mathcal{D}_n}d_j + \sum_{v\in\mathcal{V}^{\textrm{rec}}_n}\delta_v - \sum_{v\in\mathcal{V}^{\textrm{snd}}_n} \delta_v \leq \bar{d}^{\max}_n, \, (\omega^l_n, \omega^u_n) \quad n \in \mathcal{N} \label{opt:space_comp_caps} \\
&&& d \in \mathcal{C}_d, \, p \in \mathcal{C}_p, \, \theta \in \mathcal{C}_{\theta}, \, \delta \in \mathcal{C}_\delta \label{opt:space_caps}
\end{align}
\end{subequations}
where the set $\mathcal{C}_\delta := \{\delta \, | \, \delta_v \in [0, \bar{\delta}_v] \,\, \forall \,\, v \in \mathcal{V}\}$ captures the capacity constraints for virtual links. Comparing this formulation with the previous formulation \eqref{opt:disagg_vl}, we observe that the social surplus \eqref{opt:space_obj} now captures the operational cost of virtual links; moreover, the balance constraints \eqref{opt:space_balance} now include spatial virtual shifts. The dual variables associated with the computing capacity constraints \eqref{opt:space_comp_caps} are denoted $\omega^l_n$ and $\omega^u_n$, respectively. Note that the operational cost of virtual links resembles that of operational costs of physical transmission and, as the name suggest, these capture costs associated with load shifting (e.g., data transfer costs). In this market, each consumer $j\in \mathcal{D}$ is charged with the electricity price at the hub node and also remunerated by the shifting service provided through virtual links. As shown in Section \ref{sec:basic_vl_market}, the market pays off virtual links by the price difference between the sending and receiving nodes. 

\subsection{Basic Formulation with General Virtual Links (Temporal)}

The concept of virtual links naturally arises from the ability of DaCes to offer geographical load-shifting flexibility; however, this concept can also be used to capture temporal flexibility. This is key because DaCes are also able to schedule tasks over time in a way that they find most efficient/profitable. The key is to capture temporal shifting flexibility by using virtual links considering a time horizon as a linear network (with nodes defining time locations). We thus have that virtual links {\em transport} load from a given time location to another time location in the future. 

To see how to incorporate temporal virtual links in the clearing model, we consider a time horizon given by the time nodes $\mathcal{T} = \{t_1, t_2, ..., t_T\}$. For simplicity, we consider a network with a single spatial node (no transmission network is present in this setting). For each time node $t \in \mathcal{T}$, the DaCe bids a price $\alpha^d_t$ and a capacity $\bar{d}_t$ that represents the amount of load requested. Similarly, we consider a supplier that bids a price $\alpha^p_t$ and a capacity $\bar{p}_t$ at time $t\in \mathcal{T}$. The clearing formulation will find optimal levels for load satisfaction $d_t$ and generation $p_t$ for each time $t \in \mathcal{T}$. 

Each virtual link $v \in \mathcal{V}$ branches from a time node $t$ to a later time node $t'$.  The virtual link bids into the market at a  price $\alpha^\delta_v$ and capacity $\mathcal{\delta}_t$. The load at each time $t \in \mathcal{T}$ is associated with a set of virtual links $\mathcal{V}_t\subseteq \mathcal{V}$. For each $v \in \mathcal{V}_t$, we have $\textrm{snd}(v) = t$ (since each consumer can only bid flexibility going from the current time to a later time). The market clearing process will find the optimal time shift flows $\delta_v$ for each $v \in \mathcal{V}$. A distinguishing feature of temporal virtual links (compared to spatial virtual links) is that they are naturally unidirectional. The temporal model needs to establish balance constraints for each time node. The load cleared/withdrawn at $t$ is:
\begin{align}
 \hat{d}_t = d_t + \sum_{v \in \mathcal{V}^{in}_t}\delta_v - \sum_{v \in \mathcal{V}^{out}_t}\delta_v. 
\end{align} 
Interestingly, we note that this power balance is similar to that of an {\em energy storage} system.  This indicates that storage systems act as transporters/carriers of load and thus can be remunerated as flexibility providers.  This also highlights that virtual links provide a mechanism to remunerate technologies that can provide load-shifting flexibility (e.g., buildings, manufacturing, batteries). 

The clearing formulation for this setting is:
\begin{subequations}
\label{opt:temp}
\begin{align}
& \underset{d,p,\delta}{\text{min}} && \sum_{t \in \mathcal{T}} \alpha_{t}^p p_{t} + \sum_{v \in \mathcal{V}} \alpha^\delta_v \delta_v - \sum_{j\in\mathcal{D}} \alpha_{j}^d d_{j} \label{opt:temp_obj} \\ 
& {\text{s.t.}} && p_t = d_t + \sum_{v \in \mathcal{V}^{in}_t}\delta_v - \sum_{v \in \mathcal{V}^{out}_t}\delta_v, \quad t \in \mathcal{T} \label{opt:temp_balance} \\ 
&&& 0 \leq d_t + \sum_{v\in\mathcal{V}^{\textrm{rec}}_t}\delta_v - \sum_{v\in\mathcal{V}^{\textrm{snd}}_t} \delta_v \leq \bar{d}^{\max}_t, \, (\omega^l_t, \omega^u_t) \quad t \in \mathcal{T} \label{opt:temp_comp_caps} \\
&&& 0 \leq p_t \leq \bar{p}_t, \quad t \in \mathcal{T} \label{opt:temp_cap_p} \\
&&& 0 \leq d_t \leq \bar{d}_t, \quad t \in \mathcal{T} \label{opt:temp_cap_d} \\
&&& 0 \leq \delta_v \leq \bar{\delta}_v, \quad v \in \mathcal{V} \label{opt:temp_cap_vl} 
\end{align}
\end{subequations}
The load at each time $t$ is charged at the corresponding cleared price and also remunerated by the shifting service provided through virtual links; the consumer flexibility is remunerated based on the price difference between the sending and receiving times. In other words, a load shift will occur provided there is a price difference. We can see that the temporal formulation is directly analogous to the spatial formulation; as such, we can use virtual links to unify  space-time shifting. 

\begin{figure}[!htb]
\centering
\includegraphics[width=1\textwidth]{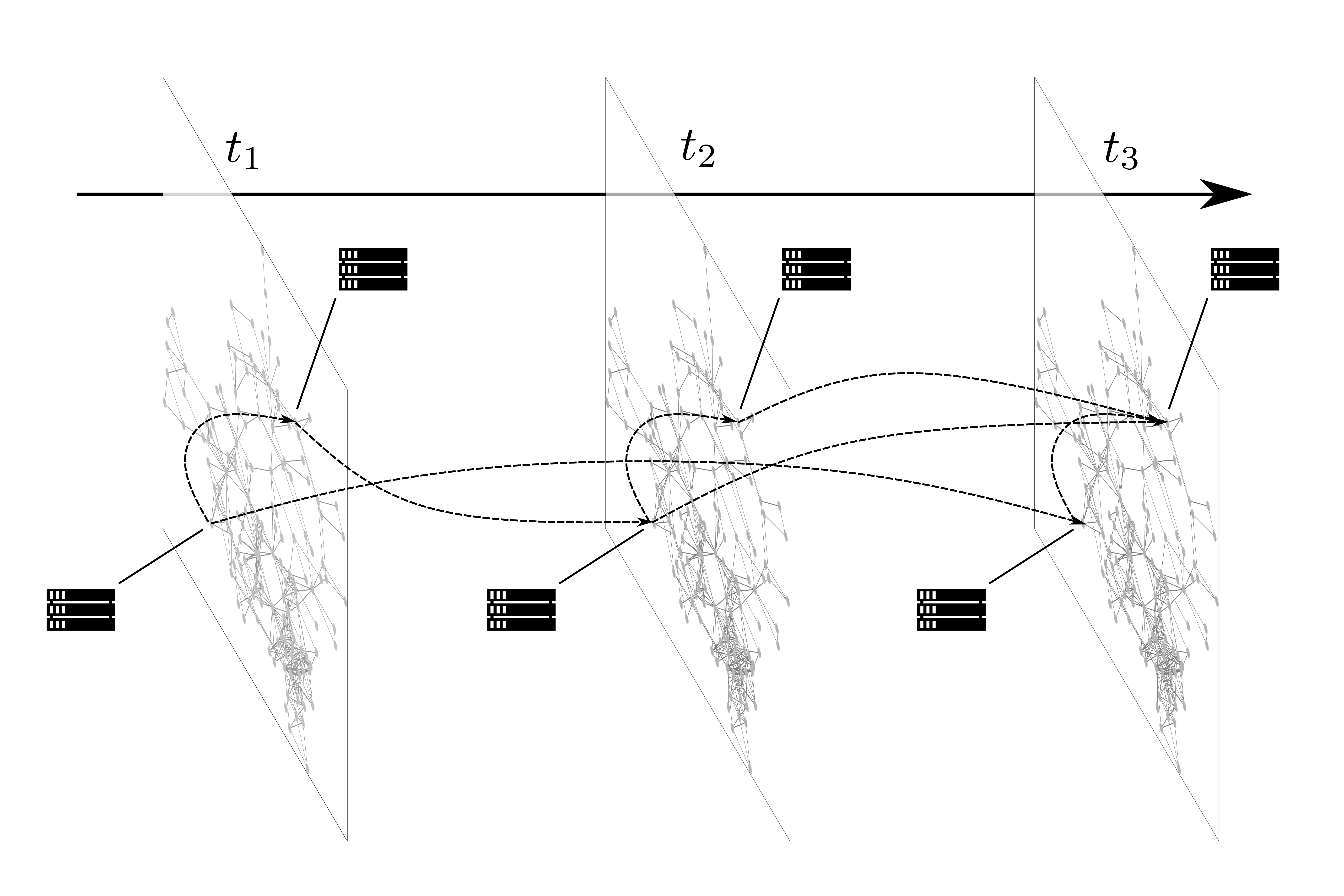}
\caption{\footnotesize Illustration of space-time load shifting using virtual links.}
\label{fig:st_model}
\end{figure}

\section{Market Formulation with Space-Time Virtual Links} \label{sec:properties}

In this section, we establish properties for a general clearing framework with space-time virtual links \eqref{opt:st}. This formulation unifies all the formulations that we have previously discussed. We will show that this clearing formulation satisfies revenue adequacy, cost recovery, and provides a competitive equilibrium. Moreover, we explore the effect of consumer flexibility on space-time price behavior; specifically, we will show that virtual links mitigate volatility.  

Central to our results is the observation that virtual links can be used to treat space and time load-shifting flexibility in a unified manner, as shown in Figure \ref{fig:st_model}. In this illustration, the spatial nodes of the network are extended into a time dimension using a set of time nodes, thus becoming a space-time network (a graph). Each time slice of the space-time graph represents the state of the network at the corresponding time.  This space-time network representation is analogous to those used in dynamic network flow models. 

Consider a space-time clearing setting with spatial nodes $\mathcal{N}$ and temporal nodes $\mathcal{T}$. A node in this space-time graph is defined as the pair $(n,t)\in \mathcal{N}\times \mathcal{T}$ (we refer to $(n,t)$ as a space-time node). Participation of suppliers, consumers and transmission services is extended to include a time dimension. Specifically, participants bid prices $\alpha^p_{i,t}$, $\alpha^d_{j,t}$, $\alpha^f_{k,t}$ and capacities $\bar{p}_{i,t}$, $\bar{d}_{j,t}$,$\bar{f}_{k,t}$ at each time $t \in \mathcal{T}$ (bids are a function of time). The cleared allocations $d_{j,t}$, $p_{i,t}$, $f_{k,t}$ and $\theta_{k,t}$ are also indexed in time. 

Virtual links connect space-time nodes; each $v \in \mathcal{V}$ is associated with a sending space-time node $\text{snd}(v) = (n_{\text{snd}(v)}, t_{\text{snd}(v)})$ and a receiving space-time node $\text{rec}(v) = (n_{\text{rec}(v)}, t_{\text{rec}(v)})$. We define $\mathcal{V}_{n,t}^\textrm{snd} := \{v\in\mathcal{V} \,|\, \textrm{snd}(v) = (n,t)\} \subseteq \mathcal{V}$, $\mathcal{V}_{n,t}^\textrm{rec} := \{v\in\mathcal{V} \,|\, \textrm{rec}(v) = (n,t)\} \subseteq \mathcal{V}$ to be the set of sending and receiving virtual links at space-time node $(n,t)$. This setting captures the special case in which $v$ is a spatial virtual link if it connects nodes at  different locations but same time ($n_{\text{snd}(v)} \neq n_{\text{rec}(v)}, t_{\text{snd}(v)} = t_{\text{rec}(v)}$). Similarly, the setting captures the special case in which $v$ is a temporal virtual link if it connects nodes at different times but at the same location ($t_{\text{snd}(v)} \neq t_{\text{rec}(v)}, n_{\text{snd}(v)} = n_{\text{rec}(v)}$). The load $j\in \mathcal{D}$ at each space-time node $(n(j),t(j))$ is associated with a set of virtual links $\mathcal{V}_j$ and we have $\mathcal{V} = \bigcup_{j\in\mathcal{D}}\mathcal{V}_j$. 

\begin{figure}[!htb]
\centering
\includegraphics[width=0.7\textwidth]{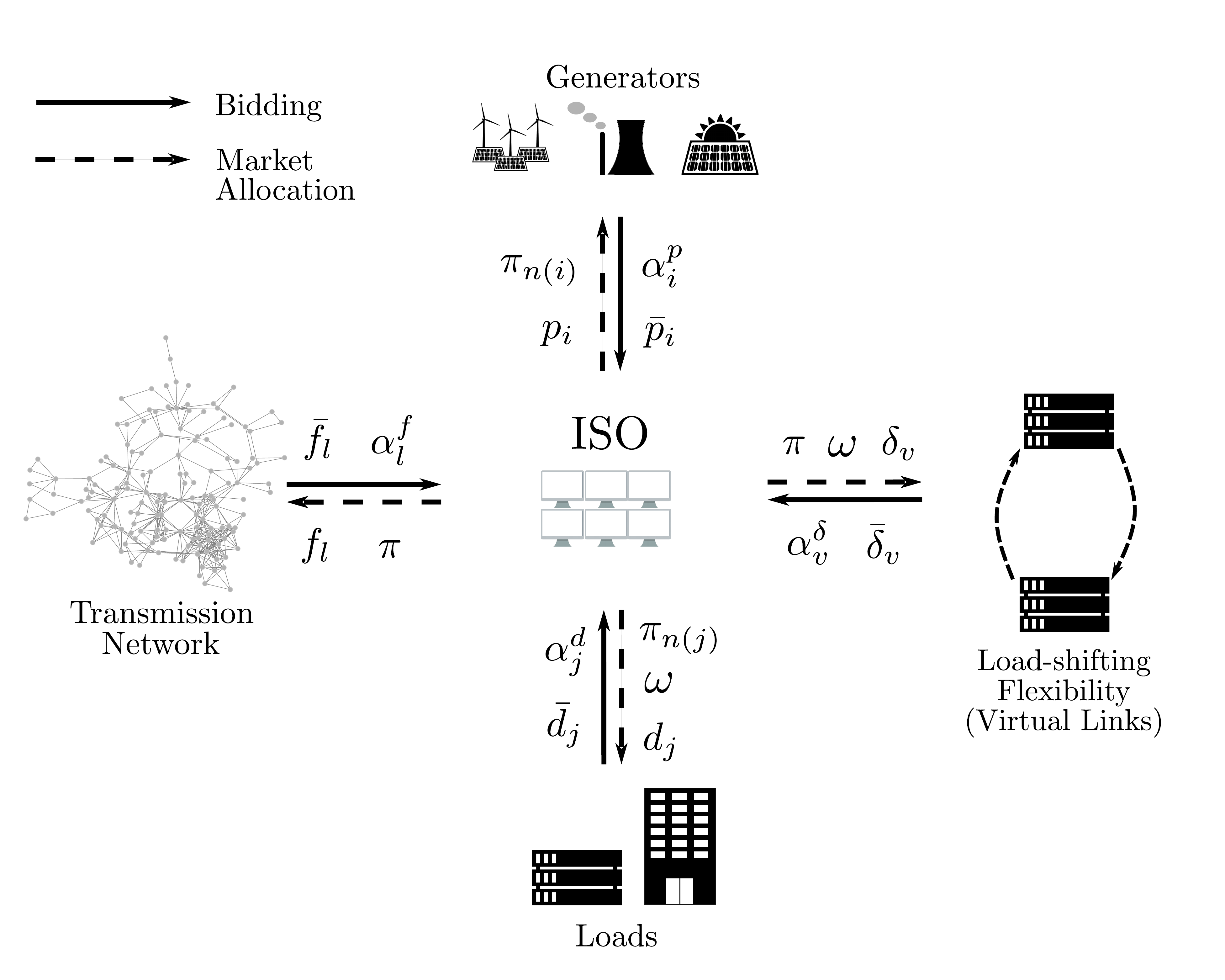}
\caption{\footnotesize Market clearing mechanism with flexible loads using space-time virtual links.}
\label{fig:space_time_infra}
\end{figure}

The clearing formulation with space-time virtual links is:
\begin{subequations}
\label{opt:st}
\begin{align}
& \underset{d, p,f,\theta, \delta}{\text{min}} & & \sum_{t \in \mathcal{T}} \Big(\sum_{i \in \mathcal{S}} \alpha_{i,t}^p p_{i,t} + \sum_{k \in \mathcal{K}} \alpha^f_{k,t} f_{k,t} - \sum_{j\in\mathcal{D}} \alpha_{j,t}^d d_{j,t} \Big) + \sum_{v \in \mathcal{V}} \alpha^\delta_{v} \delta_v \label{opt:st_obj} \\ 
& {\text{s.t.}} & &  \sum_{k \in \mathcal{K}_n^{\textrm{rec}}} f_{k,t} + \sum_{i \in \mathcal{S}_n} p_{i,t} + \sum_{v \in \mathcal{V}^{\textrm{snd}}_{n,t}} \delta_v = \sum_{k \in \mathcal{K}_n^{\textrm{snd}}} f_{k,t} + \sum_{j \in \mathcal{D}_n} d_{j,t} + \sum_{v \in \mathcal{V}^{\textrm{rec}}_{n,t}} \delta_v, \quad (\pi_{n,t}) \quad  n \in \mathcal{N}, t \in \mathcal{T} \label{opt:st_balance} \\ 
&&&  f_{l^+,t} - f_{l^-,t} = B_l(\theta_{\textrm{snd}(l),t} - \theta_{\textrm{rec}(l),t}), \quad l \in \mathcal{L}, t \in \mathcal{T} \label{opt:st_kvl} \\
&&& 0 \leq \sum_{j \in \mathcal{D}_n}d_{j,t} + \sum_{v\in\mathcal{V}^{\textrm{rec}}_{n,t}}\delta_v - \sum_{v\in\mathcal{V}^{\textrm{snd}}_{n,t}} \delta_v \leq \bar{d}^{\max}_{n,t}, \quad (\omega^l_{n,t}, \omega^u_{n,t}) \quad  n \in \mathcal{N}, t \in \mathcal{T}  \label{opt:st_comp_caps} \\
&&& (d,p,\theta,\delta) \in \mathcal{C} \label{opt:st_cap}
\end{align}
\end{subequations}
where $\mathcal{C} := \mathcal{C}^d \times \mathcal{C}^p \times \mathcal{C}^{\theta} \times \mathcal{C}^\delta$ captures the capacity constraints for all variables:
\begin{subequations}
\begin{gather}
\mathcal{C}_d := \{d\,| \, d_{j,t} \in [0, \bar{d}_{j,t}]\,\, \forall \,\, j \in \mathcal{D}, \, t \in \mathcal{T}\} \\
\mathcal{C}_p := \{p\,| \, p_{i,t} \in [0, \bar{p}_{i,t}]\,\, \forall \,\, i \in \mathcal{S}, \, t \in \mathcal{T}\} \\
\mathcal{C}_\theta := \{\theta \, | \, \theta_{\textrm{rec}(k),t} - \theta_{\textrm{snd}(k),t} \in [-\Delta\bar{\theta}_{k,t},\Delta\bar{\theta}_{k,t}] \,\, \forall \,\, k \in \mathcal{K}, t\in \mathcal{T}\} \\
\mathcal{C}_\delta := \{\delta \, | \, \delta_v \in [0, \bar{\delta}_v] \,\, \forall \,\, v \in \mathcal{V}\} 
\end{gather}
\end{subequations}
The social surplus (\ref{opt:st_obj}) captures the entire time horizon. Constraints (\ref{opt:st_balance}) are nodal balances for each space-time node (we denote the duals of these constraints as $\pi_{n,t}$). Constraints (\ref{opt:st_kvl}) are DC power flows for all times. Constraints (\ref{opt:st_comp_caps}) are computing capacity constraints for DaCes at space-time node $(n,t)$. The duals associated with these constraints are denoted $\omega^l_{n,t}\in \mathbb{R}_+$ and $\omega^u_{n,t} \in \mathbb{R}_+$.  The market clearing setting is illustrated in Figure \ref{fig:space_time_infra}. 

\subsection{Market Properties}

\begin{figure}[!htb]
\centering
\includegraphics[width=0.8\textwidth]{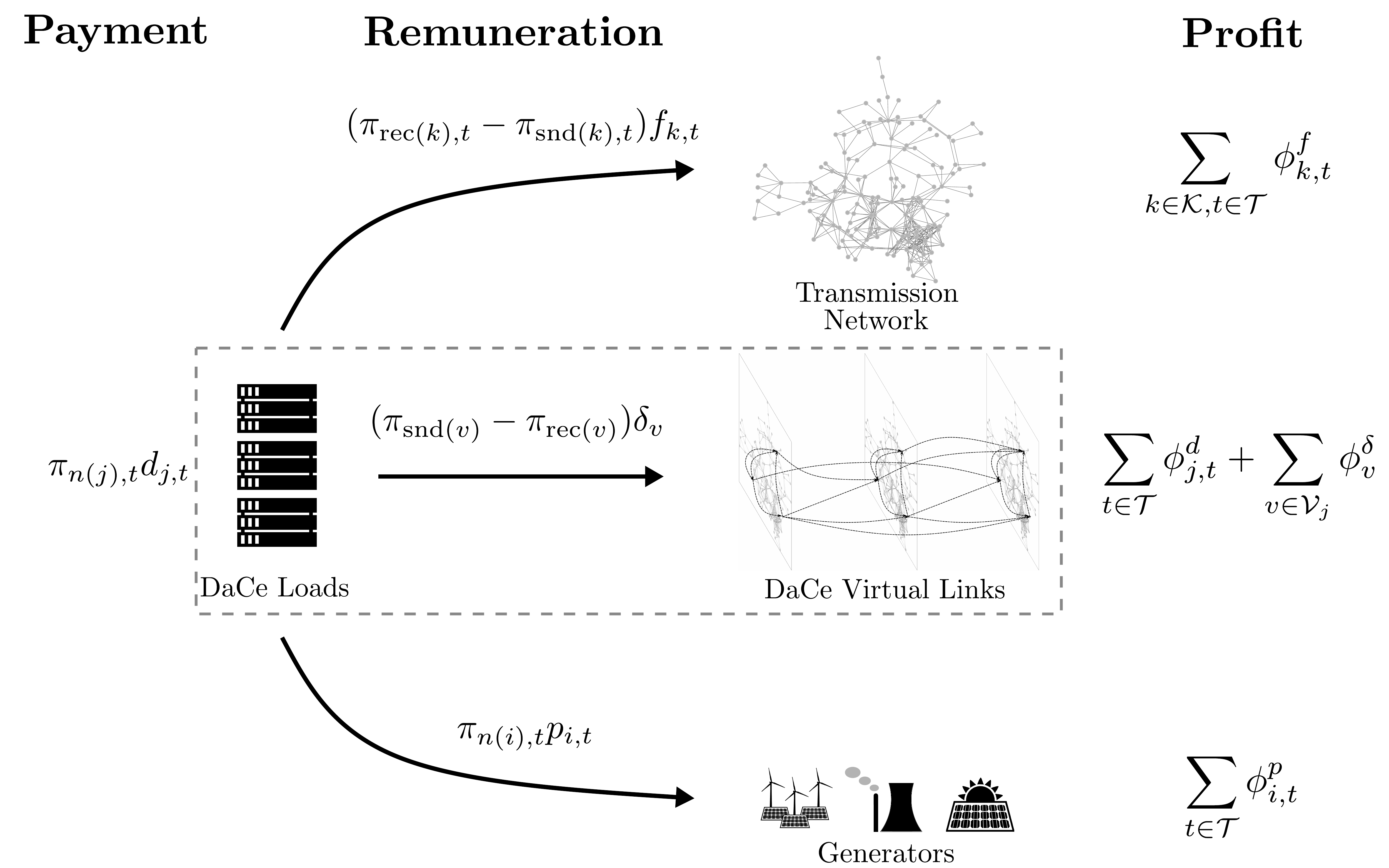}
\caption{\footnotesize Payment, remuneration, and profit of market players.}
\label{fig:remuneration}
\end{figure}

To establish market properties for \eqref{opt:st}, we formulate the partial Lagrange function of \eqref{opt:st}:
\begin{equation}
\label{eq:lagrangian}
\begin{aligned}
L(d,p,f,\theta,\delta,\pi,\omega) & = && \sum_{t \in \mathcal{T}} \Big(\sum_{i \in \mathcal{S}} \alpha_{i,t}^p p_{i,t} + \sum_{k \in \mathcal{K}} \alpha^f_{l,t} f_{l,t} - \sum_{j\in\mathcal{D}} \alpha_{j,t}^d d_{j,t}\Big) + \sum_{v \in \mathcal{V}} \alpha^\delta_v \delta_v \\
&&&- \sum_{t \in \mathcal{T}} \sum_{n \in \mathcal{N}} \pi_{n,t} \Big(\sum_{k \in \mathcal{K}_n^{\textrm{rec}}} f_{k,t} + \sum_{i \in \mathcal{S}_n} p_{i,t} + \sum_{v \in \mathcal{V}^{\textrm{snd}}_{n,t}} \delta_v - \sum_{k \in \mathcal{K}_n^{\textrm{snd}}} f_{k,t} - \sum_{j \in \mathcal{D}_n} d_{j,t} - \sum_{v \in \mathcal{V}^{\textrm{rec}}_{n,t}} \delta_v \Big) \\ 
&&& +\sum_{n\in\mathcal{N}, t\in\mathcal{T}} \omega^u_{n,t}\Big(\sum_{j \in \mathcal{D}_n}d_{j,t} + \sum_{v\in\mathcal{V}^{\textrm{rec}}_{n,t}}\delta_v - \sum_{v\in\mathcal{V}^{\textrm{snd}}_{n,t}} \delta_v - \bar{d}^{\max}_{n,t} \Big) \\
&&& -\sum_{n\in\mathcal{N}, t\in\mathcal{T}} \omega^l_{n,t} \Big(\sum_{j \in \mathcal{D}_n}d_{j,t} + \sum_{v\in\mathcal{V}^{\textrm{rec}}_{n,t}}\delta_v - \sum_{v\in\mathcal{V}^{\textrm{snd}}_{n,t}} \delta_v\Big)\\
& = && -\sum_{j\in\mathcal{D},t\in\mathcal{T}}\phi_{j,t}^d(\hat{\pi}_{n(j),t}, \alpha_{j,t}^d,d_j) - \sum_{v\in\mathcal{V}}\phi_v^\delta(\hat{\pi}_{\textrm{rec}(v)}, \hat{\pi}_{\textrm{snd}(v)}, \alpha_v^\delta, \delta_v) \\
&&& - \sum_{k \in \mathcal{K},t \in \mathcal{T}} \phi_{k,t}^f(\pi_{\textrm{rec}(k),t}, \pi_{\textrm{snd}(k),t}, \alpha^f_{k,t}, f_{k,t}) -\sum_{i \in \mathcal{S},t \in \mathcal{T}} \phi_{i,t}^p(\pi_{n(i),t}, \alpha_{i,t}^p, p_i)
\end{aligned}
\end{equation}
where we define $\omega_{n,t} := \omega^u_{n,t} - \omega^l_{n,t}$ and  $\hat{\pi}_{n,t} := \pi_{n,t} + \omega_{n,t}$. The profits for demands, virtual links, suppliers, and transmission links are (see Figure \ref{fig:remuneration}):
\begin{subequations}
\begin{align}
	\phi_{j,t}^d(\hat{\pi}_{n(j),t}, \alpha_{j,t}^d,d_j) &:= (\alpha_{j,t}^d - \hat{\pi}_{n(j),t})d_{j,t} \\
	\phi_v^\delta(\hat{\pi}_{\textrm{rec}(v)}, \hat{\pi}_{\textrm{snd}(v)}, \alpha_v^\delta, \delta_v) &:= (\hat{\pi}_{\textrm{snd}(v)} - \hat{\pi}_{\textrm{rec}(v)} - \alpha^\delta_v) \delta_v  \\
    \phi_{i,t}^p(\pi_{n(i),t}, \alpha_{i,t}^p, p_{i,t}) &:= (\pi_{n(i),t} - \alpha_{i,t}^p)p_{i,t} \\
    \phi_{k,t}^f(\pi_{\textrm{rec}(k),t}, \pi_{\textrm{snd}(k),t}, \alpha^f_{k,t}, f_{k,t}) &:= (\pi_{\textrm{rec}(k),t} - \pi_{\textrm{snd}(k),t} - \alpha^f_{k,t})f_{k,t}
\end{align}
\end{subequations} 
We can thus see that the Lagrange function \eqref{eq:lagrangian} is the negative sum of profit functions for all participants with price adjustment for DaCes due to the capacity constraints \eqref{opt:st_comp_caps}.  The presence  of capacity constraints for DaCes introduces technical difficulties in the analysis (as they couple demands and virtual links). To see this, we note that the total profit for the DaCes is: 
\begin{equation}
\begin{aligned}
 & \sum_{t\in\mathcal{T}}\left(\sum_{j\in\mathcal{D}}(\alpha_{j,t}^d - \hat{\pi}_{n(j),t})d_{j,t}\right)  + \sum_{v\in\mathcal{V}} (\hat{\pi}_{\textrm{snd}(v)} - \hat{\pi}_{\textrm{rec}(v)} - \alpha^\delta_v) \delta_v \\
= & \sum_{j\in\mathcal{D},t\in\mathcal{T}} \alpha^d_{j,t} - \sum_{v\in\mathcal{V}}\alpha^\delta_{v}\delta_v - \sum_{n\in\mathcal{N},t\in\mathcal{T}} \hat{\pi}_{n,t} \left(\sum_{j \in \mathcal{D}_n}d_{j,t} + \sum_{v\in\mathcal{V}^{\textrm{rec}}_{n,t}}\delta_v - \sum_{v\in\mathcal{V}^{\textrm{snd}}_{n,t}} \delta_v\right).
\end{aligned}
\end{equation}

The profit functions for the DaCes use $\hat{\pi} = \pi+\omega$ as prices (instead of the LMPs $\pi$). The dual variable $\omega$ adjusts the incentive for load-shifting to prevent shifting that exceeds computing capacity bounds or available loads to shift. Specifically, if $\omega_{\textrm{snd}(v)} > 0$, the upper bound of $\hat{d}_{\textrm{snd}(v)}$ is active (meaning that local loads are reaching their upper limit at $\textrm{snd}(v)$); thus, $\omega_{\textrm{snd}(v)}$ provides incentive to shift. If $\omega_{\textrm{snd}(v)} < 0$, the local loads are exactly zero, and $\omega_{\textrm{snd}(v)}$ eliminates the incentive to shift. Similarly, if $\omega_{\textrm{rec}(v)} > 0$, the upper bound of $\hat{d}_{\textrm{rec}(v)}$ is active, meaning that local loads reach the maximum at the receiving node; thus, $\omega_{\textrm{snd}(v)}$ eliminates the incentives to shift. If $\omega_{\textrm{rec}(v)} < 0$, the local loads are zero at the receiving node, and $\omega_{\textrm{rec}(v)}$ provide incentives for shifting. An alternative way of interpreting $\omega$ is as an {\em internal price factor} for DaCes; specifically, $\omega$ is the demand-supply relationship of loads within the DaCe network. If loads are not desired at a space-time node, then $\omega>0$ (the load will be pushed away from the node and therefore is not as valuable); if loads are desired at a space-time node, $\omega<0$ (the load will be attracted to the node and therefore it is valuable).  

Since we assume that strong duality holds, an optimal solution of (\ref{opt:st}) can be obtained by solving the Lagrangian dual problem:
\begin{equation}
\label{opt:dual}
    \underset{\pi, \omega}{\text{max}} \underset{(d,p,\theta,\delta) \in \mathcal{C}, f \in \mathcal{F}}{\text{min}} \, L(d,p,f,\theta,\delta,\pi,\omega)
\end{equation}
where $\mathcal{F}$ captures the DC power flow constraints \eqref{opt:st_kvl}. For fixed duals $\pi,\omega$, the Lagrange function can be decomposed into the individual profit maximization problems:
\begin{subequations}
\label{eq:max_profit}
\begin{gather}
\underset{p_{i,t} \in [0, \bar{p}_{i,t}]}{\text{max}} \phi_{i,t}^p(\pi_{n(i),t}, \alpha_{i,t}^p, p_{i,t}) \\
\underset{d_{j,t} \in [0, \bar{d}_{j,t}]}{\text{max}} \phi_{j,t}^d(\hat{\pi}_{n(j),t}, \alpha_{j,t}^d,d_{j,t}) \\
\underset{\delta_v \in [0, \bar{\delta}_v]}{\text{max}} \phi_v^\delta(\hat{\pi}_{\textrm{rec}(v)}, \hat{\pi}_{\textrm{snd}(v)}, \alpha_v^\delta, \delta_v) \\
\underset{\theta_t \in \mathcal{C}_{\theta_t},f_{k,t} \in \mathcal{F}_t}{\text{max}} \sum_{k\in\mathcal{K}} \phi_{k,t}^f(\pi_{\textrm{rec}(k),t}, \pi_{\textrm{snd}(k),t}, \alpha^f_{k,t}, f_{k,t})
\end{gather}
\end{subequations}

We now proceed to establish fundamental properties for the clearing formulation. 
\begin{theorem}
\label{thm:st_comp_eq}
The clearing formulation \eqref{opt:st} provides an allocation and prices that represent a competitive equilibrium.
\end{theorem}
\proof{}
The market is cleared by construction, because the balance constraints \eqref{opt:st_balance} are satisfied at any solution. Furthermore, \eqref{eq:max_profit} shows that the market clearing formulation \eqref{opt:st} delivers an optimal price and allocation that maximize the profit for all players. $\Box$
\endproof
\begin{theorem}
\label{thm:st_rev_ade}
The clearing formulation \eqref{opt:st} delivers an allocation and prices that satisfy revenue adequacy.
\end{theorem}
\proof{}
The following power balance holds at each space-time node:
\begin{equation*}
\sum_{k \in \mathcal{K}_n^{\textrm{rec}}} f_{k,t} + \sum_{i \in \mathcal{S}_n} p_{i,t} + \sum_{v \in \mathcal{V}^{\textrm{snd}}_{n,t}} \delta_v - \sum_{k \in \mathcal{K}_n^{\textrm{snd}}} f_{k,t} - \sum_{j \in \mathcal{D}_n} d_{j,t} - \sum_{v \in \mathcal{V}^{\textrm{rec}}_{n,t}} \delta_v = 0
\end{equation*}
Multiplying both sides by the corresponding space-time nodal price and summing over all space-time nodes, we obtain:
\begin{equation*}
\sum_{n \in \mathcal{N}, t \in \mathcal{T}} \pi_{n,t} \Big( \sum_{k \in \mathcal{K}_n^{\textrm{rec}}} f_{k,t} + \sum_{i \in \mathcal{S}_n} p_{i,t} + \sum_{v \in \mathcal{V}^{\textrm{snd}}_{n,t}} \delta_v - \sum_{k \in \mathcal{K}_n^{\textrm{snd}}} f_{k,t} - \sum_{j \in \mathcal{D}_n} d_{j,t} - \sum_{v \in \mathcal{V}^{\textrm{rec}}_{n,t}} \delta_v \Big) = 0.
\end{equation*}
This can be rewritten as:
\begin{equation*}
\begin{aligned}
& \sum_{j \in \mathcal{D}, t \in \mathcal{T}} \pi_{n(j),t} d_{j,t} & = & \sum_{i \in \mathcal{S}, t \in \mathcal{T}} \pi_{n(i),t} p_{i,t} + \sum_{n \in \mathcal{N}, t \in \mathcal{T}} \pi_{n,t} \Big( \sum_{k \in \mathcal{K}_n^{\textrm{rec}}} f_{k,t} + \sum_{v \in \mathcal{V}^{\textrm{snd}}_{n,t}} \delta_v - \sum_{k \in \mathcal{K}_n^{\textrm{snd}}} f_{k,t} - \sum_{v \in \mathcal{V}^{\textrm{rec}}_{n,t}} \delta_v \Big) \\
&& = & \sum_{i \in \mathcal{S}, t \in \mathcal{T}} \pi_{n(i),t} p_{i,t} + \sum_{k \in \mathcal{K}, t \in \mathcal{T}} (\pi_{\textrm{rec}(k),t} - \pi_{\textrm{snd}(k),t}) f_{k,t} + \sum_{v \in \mathcal{V}} (\pi_{\textrm{snd}(v)} - \pi_{\textrm{rec}(v)}) \delta_v 
\end{aligned}
\end{equation*}
The summation on the left-hand side represents the total payment by all loads, while the summations on the right-hand side represent the revenue for suppliers, transmission service provides, and virtual links (service providers), respectively. This establishes revenue adequacy. $\Box$
\endproof
\begin{theorem}
\label{thm:st_nonneg_remun}
The clearing formulation \eqref{opt:st}  delivers an allocation and prices that guarantee cost recovery for all players.  
\end{theorem}
\proof{}
Consider the allocation $(p^*,d^*,f^*,\delta_v^*)$ and duals $(\pi^*,\omega^*)$; we need to show that 
\begin{subequations}
\label{eq:nonneg_remun}
\begin{gather}
    \phi_{i,t}^p(\pi^*_{n(i),t}, \alpha_{i,t}^p, p^*_{i,t}) \geq 0 \\
	\phi_{j,t}^d(\hat{\pi}^*_{n(j),t},\alpha_{j,t}^d,d^*_j) \geq 0 \\
    \phi_v^\delta(\hat{\pi}^*_{\textrm{rec}(v)}, \hat{\pi}^*_{\textrm{snd}(v)}, \alpha_v^\delta, \delta^*_v) \geq 0 \\
    \sum_{k\in\mathcal{K}}\phi_{k,t}^f(\pi^*_{\textrm{rec}(k),t}, \pi^*_{\textrm{snd}(k),t}, \alpha^f_{k,t}, f^*_{k,t}) \geq 0.
\end{gather}
\end{subequations}
For the inner problem of \eqref{opt:dual}, $(p,d,f,\theta,\delta) = (0,0,0,0,0)$ is a feasible point (for any $(\pi, \omega)$). For fixed $(\pi, \omega)$, the inner problem is equivalent to maximizing individual profit functions in \eqref{eq:max_profit}; therefore, (\ref{eq:nonneg_remun}) hold.  $\Box$
\endproof

One can easily show that increasing load-shifting flexibility leads to a higher total social surplus. The intuition is that, with more flexibility is offered by DaCes, the ISO has more options to match demand and supplies {\em across space-time}. We use the notation $\mathcal{M}(\mathcal{V}, \bar{\delta})$ to represent the market clearing problem (\ref{opt:st}) in parametric form; the problem is a function of the set of virtual links $\mathcal{V}$ and virtual link capacities $\bar{\delta}$. We denote the corresponding optimal social surplus value as $\phi(\mathcal{V}, \bar{\delta})$.  The following result formalizes this observation.

\begin{theorem}
\label{thm:higher_welfare}
The social surplus satisfies $\phi(\mathcal{V}, \bar{\delta}) \geq \phi(\mathcal{V}^+, \bar{\delta}^+)$ if $\mathcal{V} \subseteq \mathcal{V}^+$ and $\bar{\delta}_v \leq \bar{\delta}^+_v$ for all $v \in \mathcal{V}$.
\end{theorem}

\proof{}
Let $(d,p,f,\theta, \delta)$ be a feasible solution of $\mathcal{M}(\mathcal{V}, \bar{\delta})$; the nodal balance constraints, and capacity constraints for $d, p, \theta$ remain unchanged for $\mathcal{M}(\mathcal{V}^+, \bar{\delta}^+)$ and therefore are satisfied by $(d,p,f,\theta,\delta)$. The solution $\delta_v$ satisfies the virtual link capacity constraints of $\mathcal{M}(\mathcal{V}^+, \bar{\delta}^+)$ because $v \in \mathcal{V} \subseteq \mathcal{V}_+$, and $0 \leq \delta_v \leq \bar{\delta}_v \leq \bar{\delta}^+_v$. Therefore, $(d,p,f,\theta,\delta)$ is feasible for $\mathcal{M}(\mathcal{V}^+, \bar{\delta}^+)$ and $\phi(\mathcal{V^+}, \bar{\delta}^+) \leq \phi(\mathcal{V}, \bar{\delta})$.  $\Box$
\endproof


\subsection{Pricing Properties}

We now investigate how virtual links affect price behavior. We begin by showing that the nodal prices are bounded by the bid prices of cleared players. We denote $(p^*,d^*,f^*, \theta^*, \delta^*)$ and $(\pi^*,\omega^*)$ as the optimal primal-dual allocation. At each space-time node, we define the set of cleared suppliers $\mathcal{S}_{n,t}^* := \{i \in \mathcal{S}_n \,|\, p^*_{i,t} > 0\}$, and loads $\mathcal{D}_{n,t}^* := \{j \in \mathcal{D}_n \,|\, d^*_{j,t} > 0\}$.
\begin{theorem}
\label{thm:st_price_bound}
If $\mathcal{S}_{n,t}^*$ and $\mathcal{D}_{n,t}^*$ are non-empty for $(n,t) \in \mathcal{N} \times \mathcal{T}$, the optimal prices satisfy:
\begin{equation}
\max_{i\in\mathcal{S}_{n,t}^*}\alpha_{i,t}^p \leq \pi^*_{n,t} \leq  \min_{j\in\mathcal{D}_{n,t}^*} (\alpha_{j,t}^d - \omega^*_{n(j),t})
\end{equation}
\end{theorem}
\proof{}
From Theorem \ref{thm:st_nonneg_remun} we have that:
$\phi_{j,t}^d(\pi_{j(n),t}^*, \alpha_{j,t}^d, d_{j,t}^*) \geq 0, \, \phi_{i,t}^p(\pi_{i(n),t}^*, \alpha_{i,t}^p, p_{i,t}^*) \geq 0$ holds for any $j \in \mathcal{D}_{n,t}, i \in \mathcal{S}_{n,t}$; consequently,
\begin{align*}
(\alpha_{j,t}^d - \hat{\pi}^*_{n(j),t})d^*_{j,t} &\geq 0 \\
(\pi^*_{n(i),t} - \alpha_{i,t}^p)p^*_{i,t} &\geq 0.
\end{align*}
For any $j \in \mathcal{D}_{n,t}^*$ we have that $d^*_{j,t} > 0$;  we thus have $\alpha_{j,t}^d - \omega^*_{n(j),t}\geq \pi^*_{n,t}$. Similarly, $\pi^*_{n,t} \geq \alpha_{i,t}^p$ for any $i \in \mathcal{S}_{n,t}^*$ and thus:
$$\max_{i\in\mathcal{S}_n^*}\alpha_{i,t}^p \leq \pi^*_{n,t} \leq \min_{j\in\mathcal{D}_n^*}(\alpha_{j,t}^d- \omega^*_{n(j),t})$$ 
\endproof
This result shows that cleared suppliers and consumers define the bounds for the LMP values. On the load side, we see that the price bound is on the adjusted price (effect of the shifting capacity constrains). If $\omega^*_{n(j),t}>0$ we have that the load is not desired at the node and therefore its value $\alpha_{j,t}^d- \omega^*_{n(j),t}$ is decreased; if $\omega^*_{n(j),t}<0$ we have that the load is desired at the node and therefore its value $\alpha_{j,t}^d- \omega^*_{n(j),t}$ is increased. If the net dual If the computing capacity constraints are not active, the LMPs are bounded by the load bid prices. 
\\

From cost recovery for virtual links, we have $(\hat{\pi}_{\textrm{snd}(v)}^* - \hat{\pi}_{\textrm{rec}(v)}^* - \alpha^\delta_v) \delta_v^* \geq 0$ and thus: 
\begin{equation}
\delta_{v}^* > 0 \;\;\Rightarrow\;\;  \hat{\pi}_{\textrm{snd}(v)}^* - \hat{\pi}_{\textrm{rec}(v)}^* \geq \alpha^\delta_v.
\end{equation}
This indicates that a virtual link is used only if the price difference between the receiving node and sending node is high enough to overcome its load-shifting cost (bid price). On the other hand, if the price difference is lower than the bid price, the virtual link will not be used. In short, the virtual link bid price (shifting cost) defines the minimum incentive to activate virtual links.
\\

We have shown that each virtual link $v\in \mathcal{V}$ solves the problem 
\begin{align}
\underset{\delta_v \in [0, \bar{\delta}_v]}{\text{max}} \phi_{v}^\delta(\pi_{\textrm{rec}(v)}, \pi_{\textrm{snd}(v)}, \alpha^\delta_{v}, \delta_{v}).
\end{align}
The optimal solution of this problem satisfies:
\begin{subequations}
\begin{gather}
\hat{\pi}_{\textrm{snd}(v)}^* - \hat{\pi}_{\textrm{rec}(v)}^* > \alpha^\delta_v \quad \Rightarrow\quad  \delta_v^* = \bar{\delta}_v \\ 
\delta_v^* \in (0,\bar{\delta}_v) \quad \Rightarrow \quad \hat{\pi}_{\textrm{snd}(v)}^* - \hat{\pi}_{\textrm{rec}(v)}^* = \alpha^\delta_v
\end{gather}
\end{subequations}
These results are analogous to congestion (friction) behavior observed in physical transmission networks. Specifically, the price difference between the supporting nodes of a virtual link equals the shifting cost when the virtual flow has not hit is capacity bound; this implies that, when the shift cost is zero, the prices of the supporting nodes will be the same (this helps homogenize LMPs). On the other hand, when the virtual flow hits it capacity limit, the price difference is bounded by the shift cost; since the shift cost is non-negative, we can see that price at the receiving node will be less than (or equal) that of the sending node. In other words, the receiving node cannot be higher (otherwise there is no incentive to shift load). 

A key difference between virtual flows and physical flows is that the former are not subject to any DC network constraints; as such, the only source of congestion for the virtual links is their capacity constraint. Note also that virtual flows can travel in space-time; while physical flows can only travel in space; as such, virtual flows can be used to mitigate space-time price variability. 

We note that DaCes receive the adjusted prices $\pi + \omega$ (due to the presence of computing capacity constraints); the duals $\omega$ thus play a key role that we now explain. Because $\omega_{n,t}^u\cdot \omega_{n,t}^l = 0$ for any space-time node $n,t$ (they are complementary), we have the following interpretation for possible values of $\omega$:
\begin{itemize}
    \item $\omega_{n,t} = 0$: $\omega^u_{n,t} = 0$ and $\omega^l_{n,t} = 0$. The incentive for submitting or shifting a load into space-time node $(n,t)$ is dependent on the LMPs $\pi$ alone.
    \item $\omega_{n,t} > 0$: $\omega^u_{n,t} > 0$ and $\omega^l_{n,t} = 0$. The upper bound is active, which means the computing resource is scarce at $(n,t)$. Loads submitted at and shifted into $(n,t)$ will compete for this scarce computing resource. On the other hand, virtual links flowing outward are incentivized to shift more load. 
    \item $\omega_{n,t} < 0$: $\omega^u_{n,t} = 0$ and $\omega^l_{n,t} < 0$. The lower bound is active, which means no loads are physically cleared at $(n,t)$. The DaCe thus have a higher incentive to submit loads at $(n,t)$, and virtual links have a higher incentive to shift loads into $(n,t)$. On the other hand, virtual links shifting out will compete for loads to shift.
\end{itemize}

We now explore the effect of increasing DaCe flexibility on the clearing outcomes. For simplicity, we write the Lagrangian dual problem as:
\begin{equation}
    \underset{\pi,\omega}{\text{max}} \quad \mathcal{D}(\pi,\omega), 
\end{equation}
where:
\begin{equation}
\mathcal{D}(\pi,\omega) := \underset{(d,p,\theta,\delta) \in \mathcal{C}, f \in \mathcal{F}}{\text{min}} \, L(d,p,f,\theta,\delta,\pi,\omega)
\end{equation}
To establish properties that describe the impact of adding virtual link capacity on the prices, we inspect what happens to the solution of the clearing problem if we increase the capacity of one virtual link $v \in \mathcal{V}$ by some amount $\epsilon > 0$. The capacity of $v$ is expressed as $\bar{\delta}_v = \bar{\delta}_v^0 + \epsilon$, where $\bar{\delta}_v^0$ is the original (base) capacity. We denote the Lagrangian dual problem with capacity $\bar{\delta}_v = \bar{\delta}_v^0 + \epsilon$ as:
\begin{equation}
\max_{\pi,\omega} \mathcal{D}^{\epsilon} (\pi, \omega).
\end{equation}
 We refer to this problem as $\mathcal{M}(\epsilon)$ and denote a primal-dual solution as $(p^{*\epsilon},d^{*\epsilon},f^{*\epsilon}, \theta^{*\epsilon},\delta^{*\epsilon},\pi^{*\epsilon},\omega^{*\epsilon})$. Note that $(p^{*0},d^{*0},f^{*0},\theta^{*0},\delta^{*0},\pi^{*0},\omega^{*0})$ is an optimal solution of the base problem $\mathcal{M}(0)$.

We proceed to analyze the effect of incorporating additional flexibility of DaCes. We begin with the case of adding capacity to a virtual link that is not congested. Intuitively, adding capacity to such a link should not benefit the DaCe. Our analysis shows that the unit profit for the shift (price difference between its supporting nodes minus its shift cost) does not change. The following result establishes this property.

\begin{theorem}
\label{thm:no_effect}
If $\delta^{*0}_v < \bar{\delta}_v^0$ then, for any $\epsilon > 0$, we have that:
\begin{align}\label{eq:unitprofitsame}
\hat{\pi}^{*\epsilon}_{\textrm{snd}(v)} - \hat{\pi}^{*\epsilon}_{\textrm{rec}(v)} -\alpha_v^\delta=\hat{\pi}^{*0}_{\textrm{snd}(v)} - \hat{\pi}^{*0}_{\textrm{rec}(v)} -\alpha_v^\delta.
\end{align}
\end{theorem}

\proof{Proof:}
If $\delta^{*0}_v < \bar{\delta}^0_v$, then $\hat{\pi}^{*0}_{\textrm{snd}(v)} - \hat{\pi}^{*0}_{\textrm{rec}(v)} \leq \alpha^\delta_v$, and thus:
\begin{equation*}
\phi_v^{\delta*0} = (\hat{\pi}^{*0}_{\textrm{snd}(v)} - \hat{\pi}^{*0}_{\textrm{rec}(v)} - \alpha^\delta_v)\delta^{*0}_v = 0.
\end{equation*}
This means $\mathcal{D}^{\epsilon}(\pi^{*0},\omega^{*0}) = \mathcal{D}^{0}(\pi^{*0},\omega^{*0})$ since all other profit values remain unchanged, which implies that $(p^{*0},d^{*0},\delta^{*0},f^{*0},\theta^{*0})$ also solves of $\mathcal{D}^{\epsilon}(\pi^{*0},\omega^{*0})$. We now look at an arbitrary $(\pi,\omega) \neq (\pi^{*0},\omega^{*0})$. By optimality of $\mathcal{M}(0)$, we have that $\mathcal{D}^{\epsilon}(\pi^{*0},\omega^{*0}) = \mathcal{D}^{0}(\pi^{*0},\omega^{*0}) \geq \mathcal{D}^{0}(\pi,\omega)$. For any $(\pi,\omega)$, any feasible point of $\mathcal{D}^{0}(\pi,\omega)$ is feasible for $\mathcal{D}^{\epsilon}(\pi,\omega)$. This implies $\mathcal{D}^{\epsilon}(\pi,\omega) \leq \mathcal{D}^{0}(\pi,\omega) \leq \mathcal{D}^{\epsilon}(\pi^{*0},\omega^{*0})$; thus, $(p^{*0},d^{*0},f^{*0},\theta^{*0},\delta^{*0},\pi^{*0},\omega^{*0})$ solves $\mathcal{M}(\epsilon)$. This implies that \eqref{eq:unitprofitsame} holds. $\Box$
\endproof

We now focus on the more interesting case of adding capacity to a congested virtual link. Specifically, we show that the unit profit decreases with capacity.  

\begin{theorem}
\label{thm:decrease_price_diff}
If $\delta^{*0}_v = \bar{\delta}^0_v$ then, for any $\epsilon > 0$, we have that:
\begin{equation}
\hat{\pi}^{*\epsilon}_{\textrm{snd}(v)} - \hat{\pi}^{*\epsilon}_{\textrm{rec}(v)} -\alpha_v^\delta \leq \hat{\pi}^{*0}_{\textrm{snd}(v)} - \hat{\pi}^{*0}_{\textrm{rec}(v)} -\alpha_v^\delta.\label{thm_eq1}
\end{equation}
Furthermore, for any $\epsilon_2 > \epsilon_1 > 0$,
\begin{equation}
\hat{\pi}^{*\epsilon_2}_{\textrm{snd}(v)} - \hat{\pi}^{*\epsilon_2}_{\textrm{rec}(v)} -\alpha_v^\delta\leq \hat{\pi}^{*\epsilon_1}_{\textrm{snd}(v)} - \hat{\pi}^{*\epsilon_1}_{\textrm{rec}(v)} -\alpha_v^\delta\label{thm_eq2}
\end{equation}

\end{theorem}

\proof{Proof:}
If (\ref{thm_eq1}) holds, then (\ref{thm_eq2}) holds by setting $\bar{\delta}_v = \bar{\delta}^0_v + \epsilon_1$ as the base case, and setting $\epsilon = \epsilon_2 - \epsilon_1$. We now prove (\ref{thm_eq1});  let $\Delta^*_v := \hat{\pi}^{*0}_{\textrm{snd}(v)} - \hat{\pi}^{*0}_{\textrm{rec}(v)} - \alpha^\delta_v > 0$ be the unit profit of virtual link $v$ in the base solution. Assume that $(\pi,\omega)$ satisfies $\Delta_v(\hat{\pi}) > \Delta^*_v$, where $\Delta_v(\hat{\pi}) := \hat{\pi}_{\textrm{snd}(v)}  - \hat{\pi}_{\textrm{rec}(v)} - \alpha^\delta_v$ is the unit profit of $v$ at price $\hat{\pi}$. We show that $(\pi,\omega)$ is not optimal. By optimality of $\mathcal{M}(0)$, $\mathcal{D}^{0}(\pi^{*0}, \omega^{*0}) \geq \mathcal{D}^{0}(\pi,\omega)$ for any $(\pi,\omega)$. At $(\pi^{*0}, \omega^{*0})$, $\delta_v^{*0} = \bar{\delta}_v^0 + \epsilon$ since $\Delta_v^* > 0$, and the optimal values of all other profit terms remain unchanged. Thus,
\begin{equation*}
\mathcal{D}^{0}(\pi^{*0},\omega^{*0}) - \mathcal{D}^{\epsilon}(\pi^{*0},\omega^{*0}) = \Delta_v^* (\bar{\delta}^0_v + \epsilon) - \Delta_v^*\bar{\delta}^0_v = \Delta_v^* \epsilon
\end{equation*}
The same reasoning holds for $(\pi,\omega)$ that satisfies $\Delta_v(\hat{\pi}) > 0$; we have:
\begin{equation*}
\mathcal{D}^{0}(\pi,\omega) - \mathcal{D}^{\epsilon}(\pi,\omega) = \Delta_v(\hat{\pi})\epsilon,
\end{equation*}
then we have:
\begin{equation*}
\begin{aligned}
    & \mathcal{D}^{\epsilon}(\pi^{*0},\omega^{*0}) - \mathcal{D}^{\epsilon}(\pi,\omega) & = &  (\mathcal{D}^{0}(\pi,\omega) - \mathcal{D}^{\epsilon}(\pi,\omega)) - (\mathcal{D}^{0}(\pi^{*0},\omega^{*0}) - \mathcal{D}^{\epsilon}(\pi^{*0},\omega^{*0})) + (\mathcal{D}^{0}(\pi^{*0},\omega^{*0}) - \mathcal{D}^{0}(\pi,\omega))\\
    & & = & \Delta_v(\hat{\pi})\epsilon - \Delta_v^*\epsilon + \mathcal{D}^{0}(\pi^{*0},\omega^{*0}) - \mathcal{D}^{0}(\pi,\omega) \\
    & & = & (\Delta_v(\hat{\pi}) - \Delta_v^*)\epsilon + \mathcal{D}^{0}(\pi^{*0},\omega^{*0}) - \mathcal{D}^{0}(\pi,\omega) \\
    & & > & 0
\end{aligned}
\end{equation*}
where the last inequality holds because $\Delta_v(\hat{\pi}) > \Delta_v^*$, $\mathcal{D}^{0}(\pi^{*0},\omega^{*0}) \geq \mathcal{D}^{0}(\pi,\omega)$, and $(\pi^{*0},\omega^{*0})$ solves $\mathcal{M}(0)$. Since this holds for arbitrary $(\pi,\omega)$ such that $\Delta_v(\hat{\pi}) > \Delta_v^*$, we conclude that:
\begin{equation*}
\hat{\pi}^{*\epsilon}_{\textrm{snd}(v)} - \hat{\pi}^{*\epsilon}_{\textrm{rec}(v)} \leq \hat{\pi}^{*0}_{\textrm{snd}(v)} - \hat{\pi}^{*0}_{\textrm{rec}(v)},
 \end{equation*}
 which implies \eqref{thm_eq1}. $\Box$

\endproof

Theorem \ref{thm:decrease_price_diff} indicates that increasing virtual link capacity has the effect of reducing the price difference between space-time nodes that support the virtual links. We note, however, that the ability of virtual links to reduce price volatility might be affected by computing capacity constraints (as the duals $\omega$ might distort the prices in a manner that is difficult to predict). Moreover, we note that Theorem \ref{thm:decrease_price_diff} does not guarantee convergence of the price difference to a specific value.  To address these issues, we now proceed to show that, when the capacity of a virtual link is sufficiently large, the price difference between the supporting nodes is bounded by the link shift cost. As such, the price difference can be made arbitrarily small as the shift cost is made arbitrarily small.  In order to study this convergence behavior, we apply subgradient analysis to problem \eqref{opt:dual}. A brief review of subgradient analysis and nonsmooth optimization is provided in the Appendix \ref{sec:nonsmooth}. 

We recall that minimizing the Lagrange function with respect to allocation variables is equivalent to individual profit maximization; we exploit this property to write out the optimal profit of each player as a function of $(\pi,\omega)$:
\begin{subequations}
\begin{align}
    \phi_{j,t}^{d*}(\hat{\pi}_{n(j),t}) &= \text{max}\{(\alpha_{j,t}^d - \hat{\pi}_{n(j),t})\bar{d}_{j,t}, 0\} = |\alpha_{j,t}^d - \hat{\pi}_{n(j),t}|_+\bar{d}_{j,t} \\
    \phi_{i,t}^{p*}(\pi_{n(i),t}) &= \text{max}\{(\pi_{n(i),t} - \alpha_{i,t}^p)\bar{p}_{i,t}, 0\} = |\pi_{n(i),t} - \alpha_{i,t}^p|_+\bar{p}_{i,t} \\
    \phi_v^{\delta*}(\hat{\pi}_{\textrm{snd}(v)}, \hat{\pi}_{\textrm{rec}(v)}) &= \text{max}\{(\hat{\pi}_{\textrm{snd}(v)} - \hat{\pi}_{\textrm{rec}(v)} - \alpha_v^\delta)\bar{\delta}_v,0\} = |\hat{\pi}_{\textrm{snd}(v)} - \hat{\pi}_{\textrm{rec}(v)} - \alpha_v^\delta|_+\bar{\delta}_v \\ 
    \phi_{t}^{f*}(\pi) &= \text{max}_{f_t\in\mathcal{F}_t} \sum_{k\in\mathcal{K}}(\pi_{\textrm{rec}(k),t} - \pi_{\textrm{snd}(k),t} - \alpha^f_{k,t}) f_{k,t} \label{eq:optimal_profit_flows}
\end{align}
\end{subequations}
where $|\cdot|_+ := \max\{\cdot,0\}$.  The Lagrangian dual function is thus: 
\begin{align}
\mathcal{D}(\pi,\omega) = -\sum_{t \in \mathcal{T}} \Big( \sum_{j \in \mathcal{D}} \phi_{j,t}^{d*} + \sum_{i \in \mathcal{S}} \phi_{i,t}^{p*} + \phi^{f*}_{t} \Big) - \sum_{v \in \mathcal{V}} \phi_v^{\delta*}.
\end{align}
Using linearity of subdifferential operator, we calculate the subgradient of this function with respect to each element of an arbitrary price $\pi$ as follows:
\begin{equation*}
\partial_{\pi_{n,t}}\mathcal{D} = -\left(\sum_{j \in \mathcal{D}_n} \partial_{\pi_{n,t}}\phi_{j,t}^{d*} + \sum_{i \in \mathcal{S}_n} \partial_{\pi_{n,t}}\phi_{i,t}^{p*} + \sum_{v \in \mathcal{V}^{\textrm{rec}}_{n,t} \cup \mathcal{V}^{\textrm{snd}}_{n,t} } \partial_{\pi_{n,t}}\phi_v^{\delta*} + \partial_{\pi_{n,t}} \phi_{t}^{f*}\right).
\end{equation*}
Here, $+$ denotes the Minkowski sum and the individual subgradient terms are:
\begin{equation*}
\partial_{\pi_{n(i),t}}\phi_{i,t}^{p*} = 
\begin{cases}
    \{0\}, \hat{\pi}_{n(i)} < \alpha_i^p \\
    \{\bar{p}_i\}, \hat{\pi}_{n(i)} > \alpha_i^p \\
    [0,\bar{p}_i], \hat{\pi}_{n(i)} = \alpha_i^p
\end{cases}
\end{equation*}

\begin{equation*}
\partial_{\pi_{n(j),t}}\phi_{j,t}^{d*} = 
\begin{cases}
    \{-\bar{d}_j\}, \pi_{n(j)} < \alpha_j^d \\
    \{0\}, \pi_{n(j)} > \alpha_j^d \\
    [-\bar{d}_j,0], \pi_{n(j)} = \alpha_j^d
\end{cases}
\end{equation*}

\begin{equation*}
\partial_{\pi_{\textrm{rec}(v)}}\phi_v^{\delta*} = 
\begin{cases}
    \{0\}, \, \hat{\pi}_{\textrm{snd}(v)} - \hat{\pi}_{\textrm{rec}(v)} - \alpha_v^\delta < 0\\
    \{-\bar{\delta}_v\}, \, \hat{\pi}_{\textrm{snd}(v)} - \hat{\pi}_{\textrm{rec}(v)} - \alpha_v^\delta > 0 \\
    [-\bar{\delta}_v,0], \, \hat{\pi}_{\textrm{snd}(v)} - \hat{\pi}_{\textrm{rec}(v)} - \alpha_v^\delta = 0
\end{cases}
\end{equation*}

\begin{equation*}
\partial_{\pi_{\textrm{snd}(v)}}\phi_v^{\delta*} = 
\begin{cases}
    \{0\}, \, \hat{\pi}_{\textrm{snd}(v)} - \hat{\pi}_{\textrm{rec}(v)} - \alpha_v^\delta < 0 \\
    \{\bar{\delta}_v\}, \, \hat{\pi}_{\textrm{snd}(v)} - \hat{\pi}_{\textrm{rec}(v)} - \alpha_v^\delta > 0 \\
    [0,\bar{\delta}_v], \, \hat{\pi}_{\textrm{snd}(v)} - \hat{\pi}_{\textrm{rec}(v)} - \alpha_v^\delta = 0
\end{cases}
\end{equation*}

It is difficult (if not impossible) to derive an analytic form for $\partial_{\pi_{n,t}} \phi_{t}^{f*}$ (due to the presence of DC constraints). For the following analysis, however, we only need to assume that $\partial_{\pi_{n,t}} \phi_{t}^{f*}$ is bounded. We now show the effect of increasing virtual shift capacity; in short, the virtual link capacity has a critical point beyond which the difference of $\pi + \omega$ between the connected space-time nodes will be bounded by its bid price. Once the virtual link capacity reaches this critical point, additional capacity will not be used by the market.

\begin{theorem}
\label{thm:convergence}
Assume $\partial_{\pi_{n,t}} \phi_{t}^{f*}$ is bounded for any $\pi$;  then, for any $v\in\mathcal{V}$, $\exists\; M_v > 0$ such that, if $\epsilon > M_v$:
$$\hat{\pi}^{*\epsilon}_{\textrm{snd}(v)} - \hat{\pi}^{*\epsilon}_{\textrm{rec}(v)} \leq \alpha^{\delta}_v$$
\end{theorem}

\proof{Proof:}
Consider an arbitrary virtual link $v$, the optimality conditions require that:
\begin{align*}
0 &\in \partial_{\pi_{\textrm{snd}(v)}}\mathcal{D}^{\epsilon}(\pi^{*\epsilon}, \omega^{*\epsilon})\\
0 &\in \partial_{\pi_{\textrm{rec}(v)}}\mathcal{D}^{\epsilon}(\pi^{*\epsilon}, \omega^{*\epsilon})
\end{align*}

Each term in the subgradient is an interval of possibly zero length. When the capacity of $v$ increases, all terms remain constant except for the subgradient terms $\partial_{\pi_{\textrm{snd}(v)}}\phi_v^{\delta*}$ and $\partial_{\pi_{\textrm{rec}(v)}}\phi_v^{\delta*}$. This allows us to write the sum of all other terms as constant intervals, which we denote $[a^-,a^+]$ for $\partial_{\pi_{\textrm{snd}(v)}}\mathcal{D}^{\epsilon}$ and $[b^-,b^+]$ for $\partial_{\pi_{\textrm{rec}(v)}}\mathcal{D}^{\epsilon}$, respectively. Then the subgradients can be expressed as 

\begin{equation*}
   \partial_{\pi_{\textrm{snd}(v)}}\mathcal{D}^{\epsilon} = -\left([a^-,a^+] + \partial_{\pi_{\textrm{snd}(v)}}\phi_v^{\delta*}\right)
\end{equation*}
\begin{equation*}
	\partial_{\pi_{\textrm{rec}(v)}}\mathcal{D}^{\epsilon} = -\left([b^-,b^+] + \partial_{\pi_{\textrm{rec}(v)}}\phi_v^{\delta*}\right)
\end{equation*}
Let $M_v = \text{max} \{-a^- -\bar{\delta}_v^0, b^+ -\bar{\delta}_v^0,0\}$ and suppose $\epsilon > M_v$. Given arbitrary duals $(\pi,\omega)$ that satisfy $\hat{\pi}_{\textrm{snd}(v)} - \hat{\pi}_{\textrm{rec}(v)} > \alpha_v^\delta$, we show that $\pi$ does not satisfy the optimality condition for $\mathcal{M}(\epsilon)$ if $\epsilon > M_v$. If $\pi_{\textrm{snd}(v)} + \omega_{\textrm{snd}(v)} - \pi_{\textrm{rec}(v)} - \omega_{\textrm{rec}(v)} - \alpha_v^\delta > 0$, then the subgradients at the supporting nodes become:

\begin{equation*}
\begin{aligned}
\partial_{\pi_{\textrm{snd}(v)}}\mathcal{D}^{\epsilon} \subseteq [-a^+ - \bar{\delta}_v^0 - \epsilon, -a^- - \bar{\delta}_v^0 - \epsilon] \\
\partial_{\pi_{\textrm{rec}(v)}}\mathcal{D}^{\epsilon} \subseteq [ -b^+ + \bar{\delta}_v^0 + \epsilon, -b^- + \bar{\delta}_v^0 + \epsilon]
\end{aligned}
\end{equation*}
By definition of $M_v$, we have:
\begin{equation*}
\begin{aligned}
-a^- - \bar{\delta}_v^0 - \epsilon < -a^- - \bar{\delta}_v^0 +a^- + \bar{\delta}_v^0 = 0 \\
-b^+ + \bar{\delta}_v^0 + \epsilon > -b^+ + \bar{\delta}_v^0 + b^+ - \bar{\delta}^0_v = 0
\end{aligned}
\end{equation*}
This means that the lower bound of $\partial_{\pi_{\textrm{rec}(v)}}\mathcal{D}^{\epsilon}$ is strictly positive, and the upper bound of $\partial_{\pi_{\textrm{snd}(v)}}g$ is strictly negative. Therefore, $0 \notin \partial_{\pi_{\textrm{rec}(v)}}\mathcal{D}^{\epsilon},0 \notin \partial_{\pi_{\textrm{snd}(v)}}g$, which implies $(\pi, \omega)$ is not optimal. $\Box$
\endproof

Theorem \ref{thm:convergence} shows that the price difference is eventually bounded by the shift cost in the limit of high virtual link capacity.  This result is key, as it shows that virtual links can help homogenize prices (by controlling price differences). It is important to highlight that the price differences exploited by virtual links traverse space and time and thus spatial and temporal price variability can be mitigated.  The ability to control price differences across space and time is a key benefit over power transmission (which only exploits spatial price differences). Moreover, one could argue that it is easier to expand virtual link capacity (by installing more DaCes) than it is to install more transmission lines. 

A rigorous proof of price convergence is established here for the market clearing formulation \eqref{opt:st}, which is quite general but also does not account for other features encountered in practice (e.g., ramping constraints and AC power flows). A rigorous analysis of more sophisticated formulations is challenging and is left as a topic of future work.  However, we have observed empirically that similar properties are observed in more complex formulations (in the next section we illustrate how temporal virtual links mitigate price volatility introduced by ramping constraints). 


\section{Computational Studies} \label{sec:case_study}

In this section we present case studies to demonstrate the various benefits of using the virtual link paradigm for capturing DaCe flexibility. All our models were implemented in JuMP \cite{jump_2017} and were solved using \cite{gurobi_2019}. We first analyze a small-scale model to illustrate the key results and then analyze a large-scale model to show that the results and insights are scalable. All scripts needed to reproduce the results can be found in \url{https://github.com/zavalab/JuliaBox/tree/master/VirtualLinks}. 

\begin{figure}[!ht]
    \centering
    \includegraphics[width=0.8\textwidth]{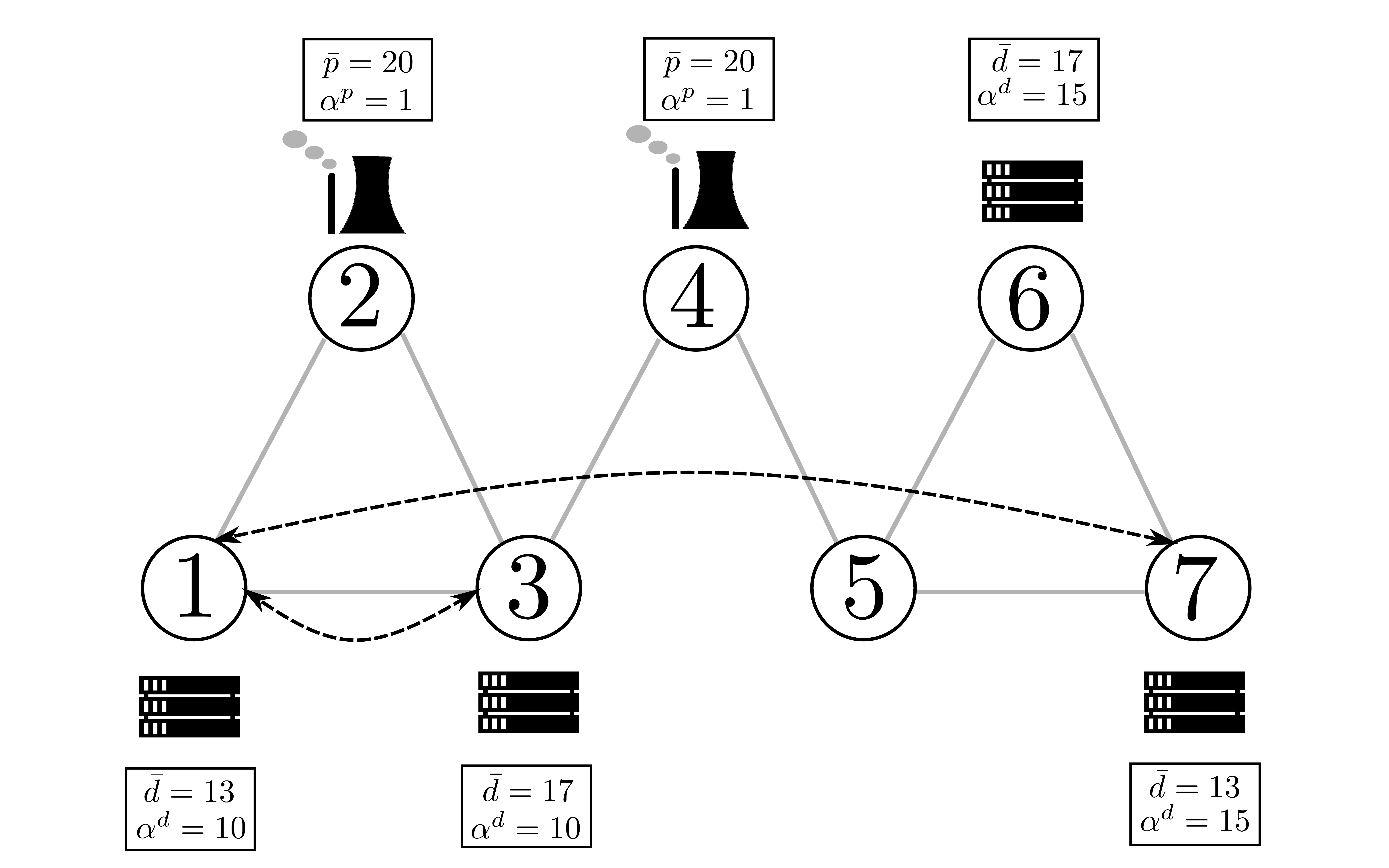}
    \caption{Scheme for 7-bus system (small generators co-located with DaCes not shown).}
    \label{fig:7bus}
\end{figure}

\subsection{7-Bus Spatial System}

We consider a 7-bus system at a fixed time, sketched in Figure \ref{fig:7bus}. Four DaCes, owned and operated by a single market player, are distributed at nodes $\{2,3,6,7\}$. Their bid prices and capacities are $\{10,10,15,15\}\,\$/\textrm{MWh}$ and $\{13,17,17,13\}\,\textrm{MWh}$, respectively. Each DaCe has a computing capacity of $20\,\textrm{MWh}$ and is co-located with a small and expensive generator with bid price $\alpha^p = 3\,\$/\textrm{MWh}$ and capacity $5\,\textrm{MWh}$. In addition, nodes 2 and 4 are connected to a large generator with bid price $\alpha^p = 1\,\$/\textrm{MWh}$ and capacity $20\,\textrm{MWh}$. The transmission network topology is highlighted using solid lines in Figure \ref{fig:7bus}. Each line has a capacity of $10\, \textrm{MWh}$ and a bid cost $0.1 \, \$/\textrm{MWh}$. 

We considered three scenarios with different virtual link capacity levels. The results are summarized in Table \ref{table:7-bus-prices}. Scenario 1 represents the base case in which no virtual links are used. Scenario 2 accounts for virtual links $(1,7)$ and $(7,1)$, both with capacity $5\,\textrm{MWh}$ and bid cost $0.3\,\$/\textrm{MWh}$. Scenario 3 is a replicate of scenario 2, except the capacity is set as $10 \, \textrm{MWh}$. Scenario 4, on top of virtual links in scenario 2, includes additional virtual links $(1,3)$ and $(3,1)$, both with capacity $5\,\textrm{MWh}$ and bid cost $0.3\,\$/\textrm{MWh}$. Scenario 5 is a replicate of scenario 4, except the capacity is set as $10 \, \textrm{MWh}$.  Scenario 6 is a replicate of scenario 5,  except that the computing capacity for each DaCe is increased from $20\,\textrm{MWh}$ to $25\,\textrm{MWh}$. Scenario 7 is a replicate of scenario 6, except that the bid costs of all virtual links are reduced to 0. 

\begin{table}
 \caption{Results for 7-bus system. Symbol $\phi$ denotes the social surplus. The surplus and dual variables are in units of USD (\$).}
 \centering
 \begin{tabular}{c | c c c c c c} 
 \hline
 Scenario & $\phi$ & $[\pi_1,\pi_2,\pi_3]$ & $\pi_4$ & $[\pi_5, \pi_6,\pi_7]$ & $[\omega_2,\omega_3,\omega_6,\omega_7]$ & $\sum d_j$ (MWh)\\
 \hline
 1 & 522 & $[3,1,2]$ & 1 & $[14.9,15,15]$ & $[0,0,0,0]$ & 50 \\ 
 2 & 577.36 & $[5,1,3]$ & 2.9 & $[14.87,15,14.93]$ & $[0,0,0,0]$ & 55 \\
 3 & 605.533 & $[10,1,5.5]$ & 1 & $[10.233,10.367,10.3]$ & $[0,0,0,0]$ & 56 \\
 4 & 582.467 & $[3,2.4,2.7]$ & 2.6 & $[14.867, 15, 14.933]$ & $[0,0,0,0]$ & 55 \\
 5 & 618.133 & $[10,1,5.5]$ & 5.4 & $[10.233,10.367,10.3]$ & $[0,4.2,0,0]$ & 57.5 \\
 6 & 639.133 & $[3.3,2.7,3]$ & 2.9 & $[3.533,3.667,3.6]$ & $[0,0,0,0]$ & 60 \\
 7 & 644.533 & $[3,1,3]$ & 1 & $[2.933, 3.067, 3]$ & $[0,0,0,0]$ & 60
\end{tabular}
 \label{table:7-bus-prices}
\end{table}

\begin{table}
 \caption{DaCe load payments and revenues for different players (in \$)}
 \centering
 \begin{tabular}{c | c | c c c c c c} 
 \hline
 Scenario & Total Load Payments & Transmission & Suppliers & Virtual Links & Total Revenue \\
 \hline
 1 & 373 & 180 & 193 & 0 & 373 \\ 
 2 & 490.13 & 181.8 & 258.67 & 49.67 & 490.13 \\
 3 & 493.63 & 273.8 & 216.83 & 3 & 493.63 \\
 4 & 459.03 & 133.8 & 264.67 & 60.57 & 459.03 \\
 5 & 508.63 & 185.8 & 306.33 & 16.5 & 508.63 \\
 6 & 203.03 & 17.8 & 179.83 & 5.4 & 203.03 \\
 7 & 181.13 & 80.8 & 100.33 & 0 & 181.13
\end{tabular}
 \label{table:7-bus-payments}
\end{table}

\begin{table}
 \caption{Profit for market players (in \$, S\# denotes supplier at node \#)}
 \centering
 \begin{tabular}{c | c c c c c} 
 \hline
 Scenario & DaCe Load & Virtual Links & Transmission & S2 & S4 \\
 \hline
 1 & 50 & 0 & 175 & 0 & 0 \\ 
 2 & 55 & 48.17 & 176.77 & 0 & 38 \\
 3 & 56 & 0 & 268.33 & 0 & 0 \\
 4 & 55 & 58.17 & 128.67 & 28 & 32 \\
 4 & 57.5 & 0 & 180.33 & 0 & 88  \\
 4 & 60 & 0 & 12.33 & 34 & 38  \\
 5 & 60 & 0 & 75.33 & 0 & 0
\end{tabular}
 \label{table:7-bus-profits}
\end{table}

Results for price behavior are summarized in Table \ref{table:7-bus-prices}. Here, $\phi$ represents the social surplus in units of USD (\$). In the base case, the LMPs show clustered patterns, where nodes in the same cycle share similar prices. We also observe a large price difference between nodes in the separate cycles. In scenarios 2 and 3, the virtual link connects across the two cycles (via node 1 and node 7) to exploit the price difference in between. We see that the price gap between node 1 and node 7 is reduced to $0.3\,\$/\textrm{MWh}$ in scenario 3, exactly the bid price of the virtual link, as predicted by the pricing properties. We also run scenario 3 with $\bar{\delta} = 1000\, \textrm{MWh}$, which gives back the same primal and dual optimal solutions except for the price at node 4 due to degeneracy. We note that the LMP at node 3 also approaches the LMPs of the other cycle even though there is no virtual link connected at node 3 yet. Similarly, the LMPs of nodes 5 and 6 come down to around $10\,\$/\textrm{MWh}$ with node 7 without virtual links directed connected. Because of the DC power flow constraints, the addition of virtual links alter the LMPs of not just the connected nodes, but also neighboring nodes in the same cluster. The values of $\omega$ for scenarios 1 and 2 are zero, meaning that computing capacity  constraints are inactive. Scenario 5 shows a case where the addition of a virtual link within cluster $\{1,2,3\}$ does not change the prices (compared with scenario 3), even if the price difference between the connected nodes is much higher than the bid cost. The reason is that computing capacity constraint is active at the destination node of the newly added virtual link (node 3) , as shown by the nonzero value of $\omega_3$. However, if the computing capacity constraints are not binding (as in scenarios 6 and 7), the price difference within the cluster $\{1,2,3\}$ becomes much smaller. In scenario 6, the price gaps across virtual links ($|\pi_1-\pi_3|$ and $|\pi_1-\pi_7|$) are exactly the bid cost of the virtual links, meaning that the price gaps converge to the best case. As an extreme case, scenario 7 shows that the prices $[\pi_1, \pi_3, \pi_7]$ become homogeneous when the bid costs are zero. These results are consistent with the pricing properties established and show how virtual links provide a mechanism to help mitigate spatial variability of prices.

Table \ref{table:7-bus-payments} summarizes results for load payments and revenues. These results verify that revenue adequacy holds for our proposed market clearing formulation. Table \ref{table:7-bus-profits} summarizes results for profits for different market stakeholders. The results verify that the clearing formulation satisfies cost recovery in all scenarios. Furthermore, we notice that virtual link profits are strictly positive and comparable to profits of load clearing in scenarios 2 and 4, but zero otherwise. This shows that virtual links provide an extra revenue stream when price volatility is high (there exist large price differences to exploit). Another interesting observation from scenarios 2 and 4 is that, when the amount of total cleared load is the same, more flexibility leads to lower load payments and higher virtual link revenue because the extra flexibility from a new virtual link provides more ways to clear the DaCe loads. This is not necessarily true when the total amount of cleared loads is different, though, because clearing more load might need to use more expensive power suppliers from the grid. When too much flexibility is provided, however, DaCes could benefit from a lower price, as shown by scenarios 6 and 7 in Table \ref{table:7-bus-payments}, but will lose the virtual link revenue streams (because of low price volatility). 

\begin{figure}[!htp]
    \centering
    \includegraphics[width=0.6\textwidth]{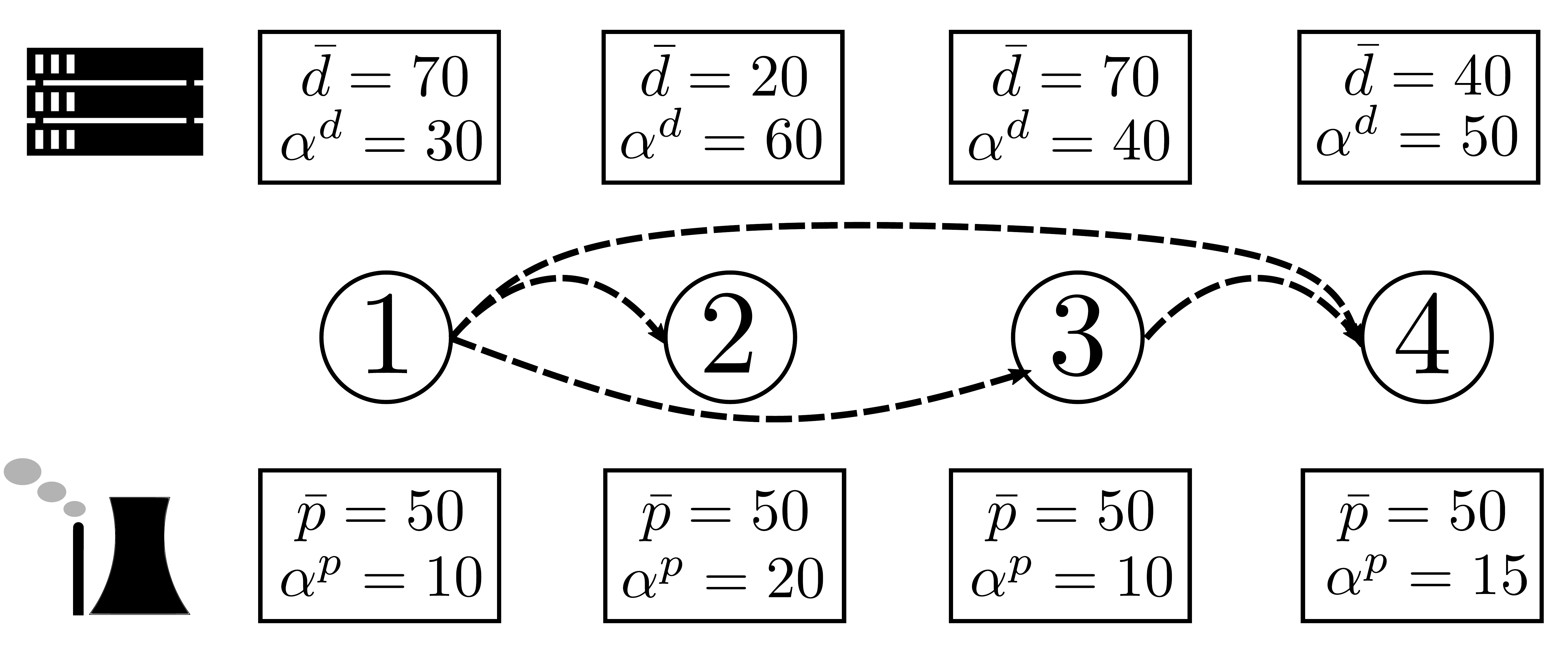}
    \caption{Scheme for 4-time system.}
    \label{fig:5times}
\end{figure}

\subsection{1-Bus Temporal System}

We now consider a single DaCe co-located with one generator over a time horizon of 4 points. The setup is a modification of the temporal case shown in  \cite{virtual_link_pscc}. The system is sketched in Figure \ref{fig:5times}. At each time interval, the DaCe can receive loads shifted from the previous time interval and delay loads to some later time interval, thus providing temporal flexibility (similar to that of a storage system). As boundary conditions, the DaCe does not receive loads at $t=t_1$, and does not delay loads at $t = 4$. The load capacity and bid costs of loads and supplies change with time. The system thus has $T=4$ time nodes and we consider 4 virtual links $\mathcal{V}:=\{(1,2), (1,3), (1,4), (3,4)\}$. The supplier capacities are set to $\bar{p}=\{50,50,50,50\}$, load capacities to $\bar{d}=\{70,25,70,40\}$, supplier bidding costs to $\alpha^p=\{10,20,10,15\}$, and load bidding prices to $\alpha^d=\{30,60,40,50\}$. We fix the bidding cost for virtual links as $\alpha^\delta=\{3,3,3,3\}$. To create extreme temporal prices differences (often seen in real systems), we also incorporate a set of ramp limit constraints $|p_{t+1} - p_t| \leq 15$. 

\begin{table*}[!h]
\centering
\caption{\small Results for one-bus network with temporal shifting flexibility. }
\begin{tabular}{c c | c c c c c} 
 \hline
 Scenario & $\bar{\delta}$ (MWh) & $\phi$ (\$) & $\pi$ (\$/MWh) & $d$ (MWh) & $p$ (MWh) & $\delta$ (MWh)\\ 
 \hline
 1 & [0,0,0,0] & 4400 & [30,-30,40,15] & [40,25,40,40] & [40,25,40,40] & [0,0,0,0] \\ 
 2 & [8,0,0,0] & 4856 & [30,-30,40,15] & [56,25,48,40] & [48,33,48,40] & [8,0,0,0] \\
 3 & [10,0,0,0] & 4970 & [30,20,40,15] & [60,25,50,40] & [50,35,50,40] & [10,0,0,0] \\
 4 & [21,0,0,0] & 5040 & [23,20,40,15] & [70,25,50,40] & [50,45,50,40] & [20,0,0,0] \\
 5 & [21,20,0,0] & 5040 & [23,20,40,15] & [70,25,50,40] & [50,45,50,40] & [20,0,0,0] \\
 6 & [11,0,11,0] & 5090 & [23,20,40,20] & [70,25,50,40] & [50,35,50,50] & [10,0,10,0] \\
 7 & [11,0,11,10] & 5197 & [30,20,40,27] & [61,25,60,40] & [50,36,50,50] & [11,0,0,10] \\
 8 & [11,0,11,20] & 5197 & [30,20,40,37] & [61,25,60,40] & [50,36,50,50] & [11,0,0,10] \\
 9 & [21,0,11,20] & 5260 & [23,20,40,37] & [70,25,60,40] & [50,45,50,50] & [20,0,0,10] \\
\end{tabular}
\label{table:5times_results}
\end{table*}

\begin{table}
 \caption{Total payments and revenue for market players (in units of \$).}
 \centering
 \begin{tabular}{c | c | c c |c c} 
 \hline
 Scenario & Load Payments & Suppliers & Virtual Links & Total Revenue \\
 \hline
 1 & 2650 & 2650 & 0 & 2650 \\ 
 2 & 3450 & 2970 & 480 & 3450 \\
 3 & 4900 & 4800 & 100 & 4900 \\
 4 & 4710 & 4650 & 60 & 4710 \\
 5 & 4710 & 4650 & 60 & 4710 \\
 6 & 4910 & 4850 & 60 & 4910 \\
 7 & 5810 & 5570 & 240 & 5810 \\
 8 & 6210 & 6070 & 140 & 6210 \\
 9 & 5990 & 5900 & 90 & 5990 \\
\end{tabular}
 \label{table:5times_payments}
\end{table}

\begin{table}
 \caption{DaCe loads and virtual link and supplier profits (in units of \$).}
 \centering
 \begin{tabular}{c | c c c} 
 \hline
 Scenario & Loads Profit & Virtual Links Profit & Suppliers Profit\\
 \hline
 1 & 3650 & 0 & 750 \\ 
 2 & 3650 & 456 & 750 \\
 3 & 2400 & 70 & 2500 \\
 4 & 2890 & 0 & 2150 \\
 5 & 2890 & 0 & 2150 \\
 6 & 2690 & 0 & 2400 \\
 7 & 1920 & 177 & 3100 \\
 8 & 1520 & 77 & 3600 \\
 9 & 2010 & 0 & 3250
\end{tabular}
 \label{table:5times_profits}
\end{table}

Nine scenarios with different temporal shifting capacities are presented in Table \ref{table:5times_results}. The results are analogous to those observed in the spatial 7-bus case and highlights how virtual links facilitate treating space and time dimensions in a unified manner. Specifically, the social surplus and the total amount of delivered loads increase with increasing shifting capacity. Price variability also becomes smaller with increasing shifting capacity. In the limit of high shifting capacity, prices converge and the differences between time nodes are bounded by the shifting cost, similar to the results of scenario 3 in the 7-bus system. Note that scenarios 1 and 2 has a negative LMP caused by the ramping limit, which is relieved by virtual links in other scenarios. Another interesting observation arises from scenarios 1 to 3, where a virtual link between $t_1$ and $t_2$ increases the amount of load cleared at $t_3$. These results show how temporal flexibility is able to relieve ramping constraints (analogous to how spatial flexibility relieves network transmission constraints). However, because temporal shifts only move in the direction of increasing time, their effects on price gaps are also unidirectional. Specifically, temporal shifts can only exploit lower prices in later times; for instance, scenarios 4 and 5 show that adding virtual link $(1,3)$ does not change the solution since the price at node 2 is higher than that at node 3. 

Table \ref{table:5times_payments} summarizes the payment and revenue results for the temporal case. We observe that revenue adequacy is satisfied for all scenarios. Table \ref{table:5times_profits} summarizes the profit results for  the temporal case. We observe that the clearing formulation satisfies cost recovery, since no participant incurs a negative profit in all scenarios. Furthermore, similar to the spatial system, the revenues and profits generated via virtual links become larger when there is more price volatility in the system.

\subsection{Space-Time IEEE-30 Bus System}

\begin{figure}[!ht]
    \centering
    \includegraphics[width=0.7\textwidth]{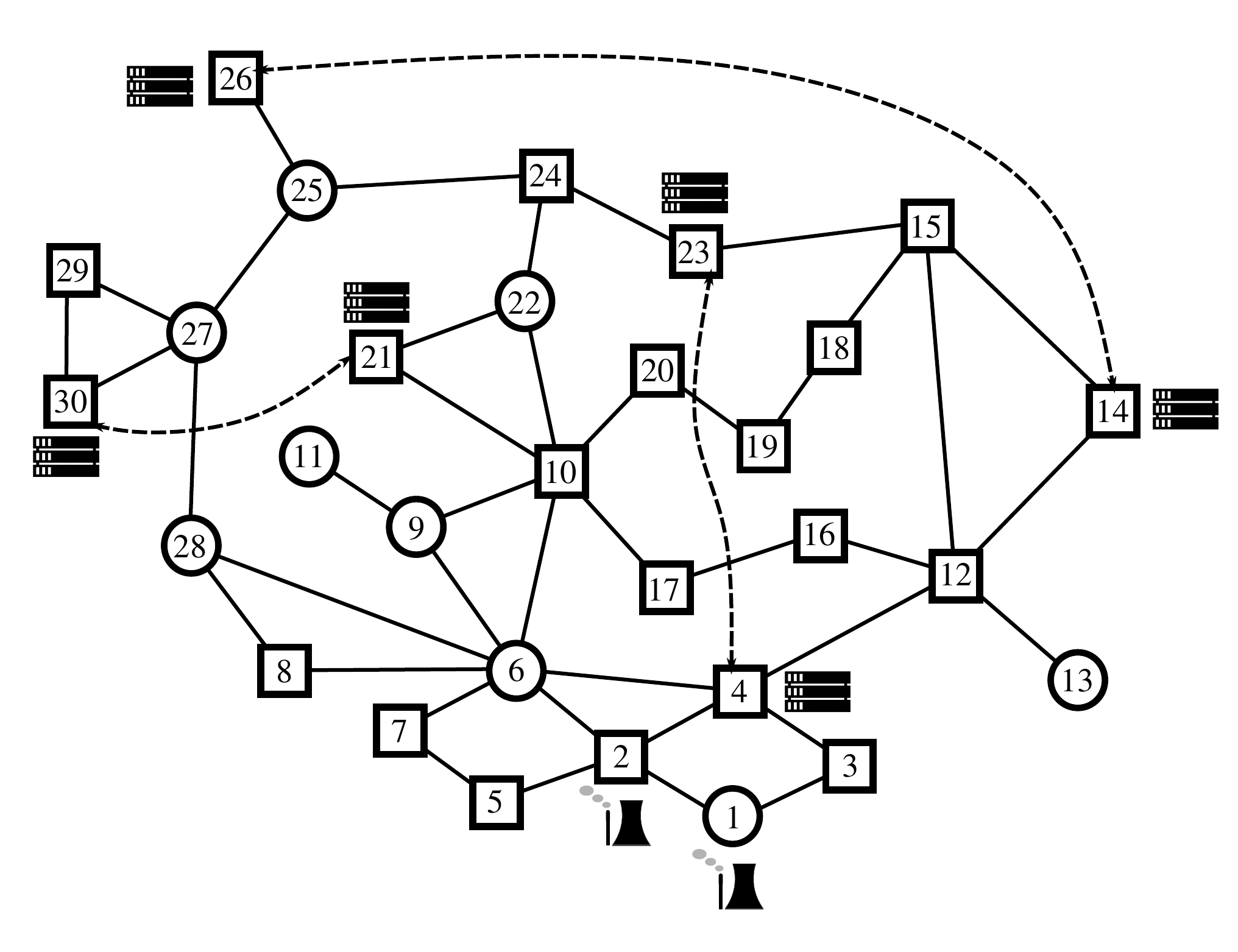}
    \caption{Scheme of IEEE 30-bus system. Squared nodes are connected to a load. Dashed curves are virtual links (not all virtual links are shown for clarity).}
    \label{fig:ieee30bus}
\end{figure}

We now consider a modified version of the IEEE 30-bus system. The network topology is presented in Figure \ref{fig:ieee30bus}. Each square node is connected to a load with varying demand capacity in time. The loads bid with the same price (200 \$/MWh) and different capacity levels. A total of six of these loads are DaCes owned by the same entity, distributed at 6 different nodes as shown in Figure \ref{fig:ieee30bus}. We run the space-time market clearing model over $T = 24$ hours. Virtual links are assigned as follows: a virtual link is assigned from one DaCes at one time, either to itself at a later time, or to another DaCe at the same time or a later time. Each virtual link has a bid price of $0\, \$/\textrm{MWh}$ and capacity of $20\,\textrm{MWh}$. The two suppliers are designated with a fixed cost and capacity over the time horizon.  

\begin{figure*}[!ht]
    \centering
	\begin{subfigure}[b]{0.49\textwidth}
         \centering
         \includegraphics[width=0.95\textwidth]{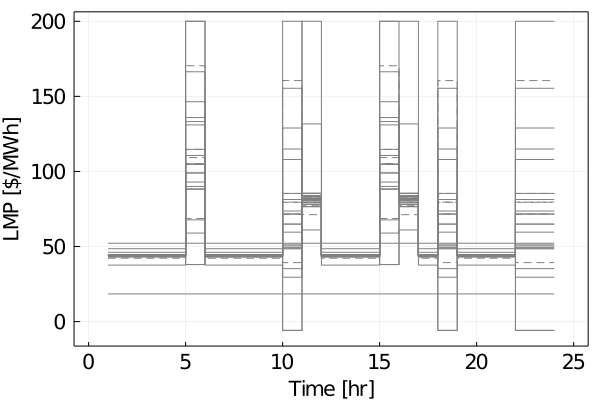}
         \caption{\small No virtual links}
         \label{fig:price_traj_novl}
    \end{subfigure}
	\begin{subfigure}[b]{0.49\textwidth}
         \centering
         \includegraphics[width=0.95\textwidth]{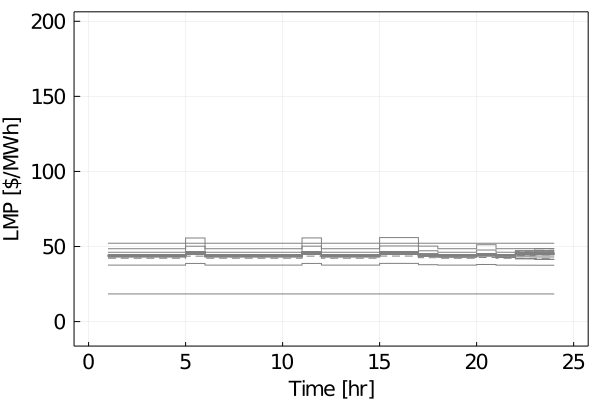}
         \caption{\small With virtual links}
         \label{fig:price_traj_vl}
     \end{subfigure}       

    \caption{Price trajectories of all buses over time. Dashed lines denote nodes with DaCes.}
    \label{fig:price_traj}
\end{figure*}

\begin{figure}[!ht]
    \centering
    \includegraphics[width=0.6\textwidth]{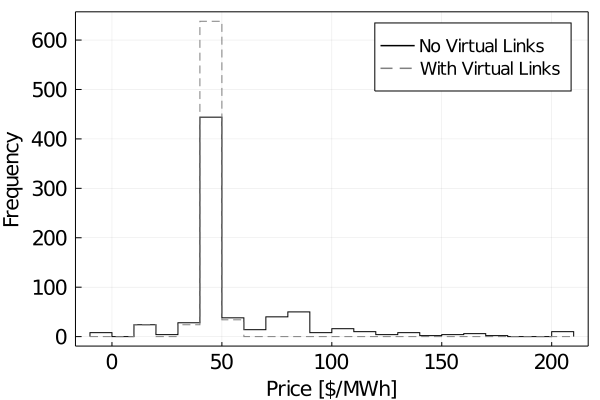}
    \caption{Space-time LMPs distribution. }
    \label{fig:LMP_histogram}
\end{figure}

The LMPs of the 30-bus system over the time horizon are plotted in Figure \ref{fig:price_traj}. We observe that virtual links are able to drastically reduce both spatial and temporal price volatility. Specifically, with no virtual links, we can observe prices reaching 200 \$/MWh in 8 out of 24 time intervals, and negative prices at 4 time intervals. Table \ref{table:stats} provide summarizing statistics for LMPs for both cases. The range, standard deviation and average deviation are all much smaller for the case with virtual links than the case with no virtual links. In addition, the case of no virtual links has a mean value that is much higher than its median value, meaning that the LMP distribution is positively skewed when there are no virtual links. The price convergence behavior can also be observed from the LMP distribution shown in Figure \ref{fig:LMP_histogram}. With virtual links, the LMPs become have less spread and exhibit a higher frequency at around 50 \$/MWh, compared to the case with no virtual links. 

\begin{table*}[ht!]
    \caption{Summarizing statistics for LMPs of IEEE 30-bus case study (in units of \$/MWh).}
    \label{table:stats}
    \begin{center}
    \begin{tabular}{c|c c}
     \hline
     LMP Statistics & No virtual links & With virtual links \\
     \hline
     Mean & 56.6 & 43.77 \\
     Median & 44.21 & 44.22 \\
     Maximum & 200.0 & 55.95 \\
     Minimum & -5.85 & 18.42 \\
     Standard Deviation & 31.36 & 5.29 \\
     Average Deviation & 21.3 & 2.32 \\
     \hline
    \end{tabular}
    \end{center}
\end{table*}

\section{Conclusions and Future Work}
\label{sec:conclusion}
We have presented a market clearing formulation to capture space-time, load-shifting flexibility provided by data centers. Load-shifting flexibility is captured using the concept of virtual links, which are non-physical pathways that can transfer power geographically and over time. We show that the proposed market clearing formulation satisfies fundamental properties (it provides a competitive equilibrium and satisfies revenue adequacy and cost recovery). Our analysis reveals that DaCes act as prosumers that are remunerated for their provision of flexibility; this remuneration is analogous to that of transmission service providers (based on nodal price differences) but is unique in that it can traverse space-time. Moreover, we show that load-shifting flexibility can help mitigate space-time price volatility; specifically, we show that prices can be made homogeneous as we increase flexibility. This new feature can be achieved because virtual links provide alternative pathways that can help relieve  physical transmission congestion. We present case studies that illustrate these effects. As part of future work, we are interested in understanding how data center flexibility could be used to mitigate risk and maximize reliability. To do so, it is necessary to develop stochastic market clearing formulations.  Moreover, we are interested in understand the effect of load-shifting flexibility on AC power flow systems and in understanding strategic bidding by DaCes that help exploit space-time price differences. 


\section*{Acknowledgments}
We acknowledge support from the U.S. National Science Foundation under award 1832208.


\bibliography{refs}

\appendix
\appendixpage
\section{Review on Non-Smooth Analysis}
\label{sec:nonsmooth}
In this section we review tools of nonsmooth analysis that we use to establish price bounding properties. This discussion is based on \cite{nonsmooth_opt}.

\subsection{Definitions}
We consider a function $f: \mathbb{R}^n \rightarrow \mathbb{R}$ that is convex but not necessarily differentiable. A vector $g \in \mathbb{R}^n$ is a subgradient of $f$ at point $x \in \mathbb{R}^n$ if for any $y \in \mathbb{R}^n$ 
\be
\label{eq:subgrad_def}
	f(y) \geq f(x) + g^T(y-x).
\ee
The right hand side of \eqref{eq:subgrad_def} is a globally valid lower bound function for $f$ that attains the exact function value at point $x$. The subdifferential $\partial f(x)$ of $f$ at $x$ is defined as the set of all subgradients at $x$:
\be
	\partial f(x) = \{g \, | \, f(y) \geq f(x) + g^T(y-x) \, \forall \, y \in \mathbb{R}^n\}.
\ee
In general, since convexity implies local Lipschitz continuity, $\partial f(x)$ is a nonempty, convex and compact set. As a special case, if $f$ is continuously differentiable at $x$, then the gradient $\nabla f(x)$ is defined at $x$ and the subdifferential becomes a singleton:
\be
	\partial f(x) = \{\nabla f(x)\}.
\ee

\subsection{Subgradient Calculus}
We now summarize a set of basic rules for subgradient calculus. If $f(x) = \sum_{i=1}^n \alpha_if_i(x)$, where $\alpha_i \geq 0$ and $f_i$ is convex. Then the subdifferential of $f$ is
\be
	\partial f(x) = \sum_{i=1}^n \alpha_i \partial f_i(x)
\ee
where the $+$ operator is the Minkowski sum operator for sets:
\be
	A + B = \{a + b \, | \, a \in A, b \in B \}
\ee
In our analysis we frequently encounter functions of the following form:
\be
	f(x) = \max_{i\in\{1,2,...n\}}\{f_i(x)\}
\ee
where each $f_i:\mathbb{R}^n \rightarrow \mathbb{R}$ is convex. The subdifferential of $f$ is
\be
	\partial f(x) = \text{conv} \bigcup_{i \in I(x)}\partial f_i(x)
\ee
where $\text{conv}(\cdot)$ is the convex hull operator and $I(x) := \{i \, | \, f_i(x) = f(x)\}$ denotes the set of active functions. When $f_i$'s are differentiable, we have
\be
	\partial f(x) = \text{conv} \{\nabla f_i(x) \,|\, i \in I(x)\}
\ee
Note that all the subgradient calculus rules reviewed above reduce to multivariate calculus if $f$ is continuously differentiable at $x$ (all $\partial f_i$ can be replaced by $\nabla f_i(x)$). 

\subsection{Optimality Conditions}
For convex $f$, $x^*$ is a global minimizer of $f$ if and only if $0 \in \partial f(x^*)$. For the case in which $f$ is continuously differentiable at $x^*$, the optimality conditions reduce to $\nabla f(x^*) = 0$, which are necessary and sufficient optimality conditions for convex minimization problems.

\end{document}